\documentclass{aa}
\usepackage{graphicx}

\begin{document}

\title{An analysis of the observed radio emission from planetary nebulae}

\author{N. Si\'odmiak \and R. Tylenda}

\institute{N. Copernicus Astronomical Center, Department for Astrophysics,
Rabia\'nska 8, 87--100 Toru\'n, Poland \\
Toru\'n Center for Astronomy, N. Copernicus University, Gagarina 11, 
87-100 Toru\'n, Poland \\ }

\abstract{
We have analysed the radio fluxes for 264 planetary nebulae for which
reliable measurements of fluxes at 1.4 and 5~GHz, and of nebular diameters
are available. For many of the investigated nebulae the optical thickness
is important, especially at 1.4~GHz. Simple models like the one specified
only by a single optical thickness or spherical, constant density shells do
not account satisfactorily for the observations. Also an $r^{-2}$ density
distribution is ruled out. A reasonable representation of the observations
can be obtained by a two-component model having regions of two different values
of optical thickness. We show that the nebular diameters smaller than
$10\arcsec$ are uncertain, particularly if they come from photographic plates
or gaussian fitting to the radio profile. While determining the interstellar
extinction from an optical to radio flux ratio caution should be paid to
optical thickness effects in the radio.
We have developed a method for estimating the value of self absorption.
At 1.4~GHz self absorption of the flux is usually important and can exceed
a factor of 10. At 5~GHz
self absorption is negligible for most of the objects although in some cases
it can reach a factor of 2. The Galactic bulge planetary nebulae
when used to calibrate the Shklovsky method give the mean nebular mass of 
$0.14\, M_{\odot}$. The statistical uncertainty of the Shklovsky distances
is smaller than factor 1.5.
}

\titlerunning{Radio emission from planetary nebulae}

\authorrunning{Si\'odmiak \& Tylenda}

\maketitle

\section{Introduction}

Planetary nebulae emit in radio wavelengths due to free-free transitions 
in the ionized gas. Numerous analyses of
individual objects and statistical studies of large samples often use the
radio flux as one of principal observational parameters.
The radio measurements, as free from
interstellar extinction, provide us with direct information on nebular
emission and can be used for determining luminosities and nebular masses.
If a sufficiently high spatial resolution is available the radio observations
can be used for studying structure of the nebulae and, in particular, 
to derive nebular dimensions.
Radio surveys are also useful for identifying new planetary nebulae,
especially in distant, heavily dust obscured regions, for instance in the 
Galactic bulge. Finally, comparison of the observed radio flux to the flux
in optical recombination lines (usually H$\beta$) is one of the
methods for measuring interstellar extinction to the planetary nebulae.

Unlike in the optical region, the planetary nebulae in the radio frequencies 
can be optically thick. This is expected to occur especially for young, dense
nebulae observed at lower frequencies. If it happens optical thickness 
effects have to be taken into account while interpreting the radio 
measurements. 

Systematic surveys of planetary nebulae have usually been done for single
frequencies. Until recently the largest data sets have been available 
for 5\,GHz.
Surveys have been done with single dish telescopes (e.g at Parkes --
Milne \& Aller 1975, Milne 1979) and with multi dish instruments (mostly
with VLA -- e.g. Zijlstra et~al. 1989, Aaquist \& Kwok 1990). Recently a 
large set of flux measurments have been published for 1.4\,GHz by 
Condon \& Kaplan (1998). The data come from the NRAO VLA Sky Survey and the 
flux measurements have been done for 680 planetary nebulae.

In this paper we present results of an analysis of the observed fluxes at
1.4 and 5\,GHz for a large sample of planetary nebulae. We have selected
those objects for which reliable flux measurements at both 
frequencies are available and for which nebular dimensions are known. In
this way we have obtained a sample of 264 planetary nebulae. With these data
we can study optical thickness effects and test simple nebular models. An
analysis of consistency of the data allows us to draw conclusions on accuracy
of the measurements. With additional observational data from the optical
region we discuss problems relevant to observational determination of 
interstellar extiction and distances.

\section{Observational data}

The observational data we have used in the present study are given in Table~1.
Column (1) containes the PN~G numbers while usual names of the planetary 
nebulae are given in column (2). 

Column (3) gives the flux measurements at 1.4\,GHz (in mJy)
from Condon \& Kaplan (1998). We have retained only those objects for which
the accuracy of the measurements, according to Condon \& Kaplan, is better
than 10\%. 

The observed fluxes at 5\,GHz (in mJy) with the references are given 
in columns (4) and (5), respectively. We have restricted our compilation to 
the results from VLA only. Our preliminary analysis
has shown that the measurements from single dishes (mostly from Parkes) do
not give consistent results for optically thin nebulae (for more discussion 
see Sect.~3.1). From the VLA results we have dismissed objects 
with uncertain (according to remarks in the original papers) flux measurements. 
We have also excluded objects with the measured 5\,GHz flux below 10 mJy (as
less reliable according to Stasi\'nska et~al. 1992) unless two
independent observations gave consistent results or the data come from 
carefull observations specially done to measure faint planetaries 
(e.g. Pottasch \& Zijlstra 1994).

Columns (6) - (10) contain data on the observed nebular dimensions.
The angular diameters (in arcsec) derived from radio observations
(in overwhelming cases at 5\,GHz) with the references are given in columns
(6) and (7), respectively. For well resolved nebulae the diameters have
been derived from the radio contours at 10\% of the peak flux
density. In other cases the data have usually come from
gaussian fitting. Some authors (e.g. Zijlstra et~al. 1989, Aaquist
\& Kwok 1990) have deconvolved the results according to the model of
Panagia \& Walmsley (1978), which essentially resulted in multiplying
the gaussian angular diameter by factor 1.8. We have done the same if only
gaussian diameters have been published in original papers.

The nebular diameters (in arcsec) obtained
from optical observations are shown in column (8). The relevent references
are given in column (9). When possible we have chosen the data from H$\alpha$
(or H$\alpha$~+~[NII]) images. In the case of objects having extended, diffuse
outer structures or haloes we have taken the diameters corresponding to the
bright main nebula. Unfortunately we have not been able to use many of 
the dimensions published in more recent CCD image catalogues 
(Schwarz et~al. 1992, Manchado et~al. 1996, G\'orny et~al. 1999). 
They corresponded to the maximum extension of the recorded nebular emission 
and it was often clear from the published images that the principal nebular 
emission came from a much smaller, central region. In many
cases the dimensions are from photographic plates and the results are
expected to be more uncertain, especially for small nebulae. ST stands for
"stellar".

The nebular diameter is a crucial parameter for our study presented in the
subsequent sections. Yet it is often an ill defined or uncertain quantity
either due to intrinsic (complex and irregular) structure of the object or
due to the way used to measure the diameter. It
would be ideally to have a quantity determined in the same way for all the
objects, i.e. from the extension of the nebular emission at a 10\% level of
the peak value. It is clear that in our compilation the data are of
different reliability. From a closer inspection of the observational
material used to determine the dimensions we have attempted to pick up more 
reliable measurements. They are marked with a + sign in columns (6) and (8).
In the case of the radio diameters these are objects for which
the published maps were of good resolution and good dynamic range.
These diameters have primarily been derived from the contours at 10\% of the 
peak flux density. For optical measurements this concerns well resolved 
nebulae, usually with a well defined outer rim.

In column (10) we give our adopted diameters. If the
radio data were consistent with the optical measurements we have simply
taken a mean value from the two. In cases of discrepant data
we have adopted values which tend to be closer to more reliable 
estimates, e.g. having assigned a + sign. For small nebulae ($< 10 \arcsec$) 
we have tended to adopt the radio measurements especially if the optical 
data came from photographic plates.

\section{Analysis of the data and theoretical modelling}

In the case of optically thin emission standard theoretical formulae 
(see e.g. Pottasch 1984) predict the ratio of the flux at 5~GHz (actually 
4.885~GHz) to that at 1.4~GHz to be 0.883. As can be seen from Table~1, for 
many objects the observed ratio is well above this value.
A straightforward interpretation is that these
planetary nebulae are not optically thin in the radio frequencies. The
optical thickness is always larger for lower frequencies. Therefore the
1.4~GHz flux is expected to be more surpressed by self absorption and the
5~GHz to 1.4~GHz flux ratio should increase with the increasing optical
thickness. In the optically thick case, when the emitted spectrum approaches 
the black body limit, the flux ratio reaches the value of 12.18.

In order to analyse the problem quatitatively we have plotted the
5~GHz to 1.4~GHz flux ratio against the radio brightness temperature,
$T_{\rm b}$, calculated from the 5~GHz flux and the adopted angular
diameter. If the flux is expressed in mJy and the diameter, $\Theta$, 
in arcsec then the brightness temperature in K can be determined from
\begin{equation}
\label{tb1}
T_{\rm b} = 73.87\, F_{\rm 5GHz}/ {{\rm \Theta}^2}.
\end{equation}
$T_{\rm b}$ can be considered as a measure of the optical
thickness. With the increasing optical thickness $T_{\rm b}$ is expected
to increase and for the optically thick limit $T_{\rm b}$ reaches the
value of the electron temperature in the nebula.

The resultant diagram is shown in Fig.~1. The points represent the observed 
positions of the objects from Table~1. The curves give the theoretical
predictions calculated from the basic model described in the following
subsection. 

\begin {figure}
 \resizebox{\hsize}{!}{\includegraphics{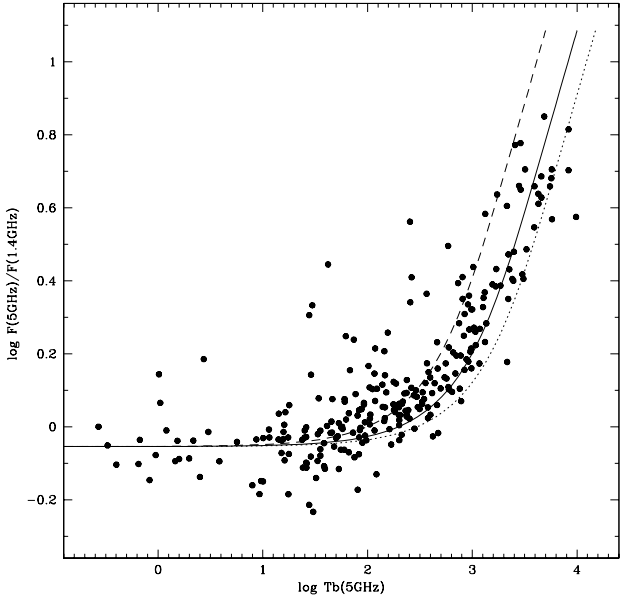}}
 \caption{The 5\,GHz to 1.4\,GHz flux ratio against the radio brightness
temperature at 5~GHz. Curves show predictions from the basic model for 
three values of $T_{\rm e}$, i.e. $0.5\,10^4$~K (dashed), $1.0\,10^4$~K (full)
and $1.5\,10^4$~K (dotted).}
\end{figure}

\subsection{Basic model}

Suppose that the nebula is an uniform 
region characterized by an electron temperature, $T_{\rm e}$, and an
optical thickness, $\tau_0$, at a reference frequency $\nu_0$. For other
frequencies the optical thickness is given as (see e.g. Pottasch 1984)
\begin{equation}
\label{tau}
          \tau_\nu = \tau_0 (\nu/\nu_0)^{-2.1}
\end{equation}
The flux, $F_\nu$, from such a model nebula
observed at a solid angle $\Omega$ can be
calculated from (see e.g. Pottasch 1984)
\begin{equation}
\label{fnu}
F_\nu = {{{2\nu^2}kT_{\rm e}}\over {c^2}}(1-e^{-\tau_\nu}) \Omega
\end{equation}
while the surface brightness temperature is given by
\begin{equation}
\label{tb2}
T_{\rm b}(\nu) = T_{\rm e}(1-e^{-\tau_\nu})
\end{equation}
The ratio of the fluxes at frequencies $\nu_1$ and $\nu_2$ is thus
\begin{equation}
\label{f}
{F_{\nu_1}\over F_{\nu_2}} = {{\nu_1^2 (1-e^{-\tau_{\nu_1}})}\over
{\nu_2^2 (1-e^{-\tau_{\nu_2}})}}
\end{equation}

The results obtained from the above model for constant values of
$T_{\rm e}$ and varying $\tau_0$ are shown as curves in Fig.~1. The three
curves coresspond to the electron temperature of $0.5\,10^4$~K, 
$1.0\,10^4$~K and $1.5\,10^4$~K.

As can be seen from Fig.~1 for low values of $T_{\rm b}$ the observed
points are scattered around the optically thin limit. For the nebulae with
log~$T_{\rm b} < 1.5$ (54 objects) the mean value of 
log~$F_{\rm 5GHz}/F_{\rm 1.4GHz}$
is $-0.044$ with a standard deviation of 0.104. The mean value agrees
with the optically thin theoretical limit. The deviation can be
considered as a typical accuracy of the observed flux ratio in our sample.

A diagram similar to that in Fig.~1 but using the 5~GHz fluxes from single
dish measurements (mostly from Parkes -- Milne \& Aller 1975, Milne 1979;
the 1.4~GHz flux from VLA -- Table~1)
shows the observed points for low $T_{\rm b}$ nebulae to be often well above
the theoretical value. For 21 objects with log~$T_{\rm b} < 1.5$
the mean log~$F_{\rm 5GHz}/F_{\rm 1.4GHz}$ is $+0.045$ with a deviation of 
0.121. The reason can be twofold. First, the single dish measurements, as 
done with a lower angular resolution than those from VLA, can be 
contaminated from nearby faint sources. Second, the VLA flux may refer only 
to the main bright nebula as VLA can be insensitive to faint, extended
outer nebular regions or halos. In any case the single dish measurements at
5~GHz give inconsistent results with the VLA data at 1.4~GHz and therefore
we have limited our analysis to the 5~GHz fluxes from VLA only.

Returning now to the discussion of Fig.~1 we see that with 
the increasing $T_{\rm b}$ the observed flux ratio increases in a
general agreement with the model curves. Thus the conclusion is that indeed
many planetary nebulae in our sample are not optically thin, at least
at 1.4~GHz.

As can be seen from the model curves in Fig.~1 the position in this diagram
depends not only on the optical thickness but also on the electron 
temperature in the nebula. In fact a simplistic interpretation of Fig.~1 
could be that many planetary nebulae have very low $T_{\rm e}$, i.e. well 
below $0.5\,10^4$~K, and that the most
opaque nebulae tend to have high $T_{\rm e}$. However, measurements of 
$T_{\rm e}$ from optical forbidden line ratios (e.g. Kaler 1986) virtually
do not give values as low as
$0.5\,10^4$~K. On the other hand, the most opaque nebulae are certainly 
the youngest ones and
are presumably ionized by relatively cool central stars. Thus they are
expected to have low $T_{\rm e}$.

In order to avoid ambiguities due to $T_{\rm e}$ and to concentrate 
the discussion on the optical thickness problems, later on in this paper,
we use the ratio $T_{\rm b}/T_{\rm e}$ instead of $T_{\rm b}$. As can be
seen from Eq.~(\ref{tb2}) this ratio depends on the optical thickness only.

Obviously for the observed objects we need estimates of their $T_{\rm e}$.
Unfortunately direct measurements of $T_{\rm e}$, e.g. from forbidden line
ratios, are available for a limited number of objects only. However, as
shown in Kaler (1986) the electron temperature measured from the [OIII] line
ratio is well correlated with the line intensity of 
HeII$\lambda 4686$\AA. In this way, using a compilation of the 
HeII$\lambda 4686$\AA~line intensity from Tylenda et~al. (1994),
we have been able to estimate $T_{\rm e}$ for the planetary nebulae in our
sample. For 35 objects, however, there have been no measurements of 
the HeII$\lambda 4686$\AA~line. In these cases we have adopted the 
canonical value of $T_{\rm e}$
for planetary nebulae, i.e. $1.0\,10^4$~K.

A diagram of the 5~GHz to 1.4~GHz flux ratio plotted against 
$T_{\rm b}/T_{\rm e}$ is shown in Fig.~2. The full curve represents the
results of our basic model as described above. Other curves show other
models described below.

\begin {figure}
 \resizebox{\hsize}{!}{\includegraphics{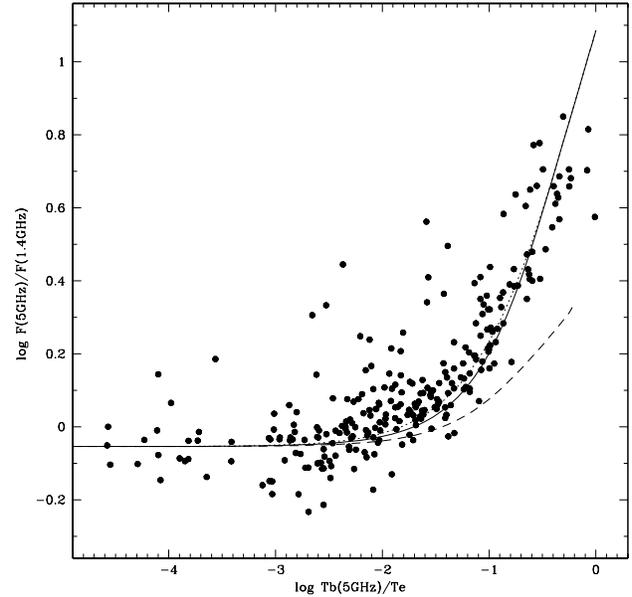}}
 \caption{The 5\,GHz to 1.4\,GHz flux ratio against $T_{\rm b}/T_{\rm e}$.
Solid line: basic model, dotted line: 
spherically symmetric shell with $R_{\rm in} / R_{\rm out} =
0.95$, dashed line: $r^{-2}$ density distribution.}
\end{figure}

As can be seen from Fig.~2 the observed positions tend to be shifted to the
upper-left in comparison to our basic model (full curve). The effect is best
seen for the objects with $-2.5 \la \mbox{log} T_{\rm b}/T_{\rm e} \la
-1.0$. The discussed model is very simple and can be considered as a very
crude representation of real nebulae. Therefore we have attempted other
models supposed to be more realistic.

\subsection{Spherically symmetric shell}

First we have analysed a spherically symmetric model in which the
absorption coefficient is constant between an inner and an outer
radius.
The flux and the surface brightness temperature can be calculated 
from numerical
integration of the radiation transfer equation.
The results, i.e. the flux ratio and $T_{\rm b}/T_{\rm e}$, for a particular
model nebula depend on two
parameters, namely the ratio of the inner to outer radius and the
optical thickness of the nebula across its centre.
It appears, however, that the results from this model
are very much the same as those from the basic model, especially for
largely filled envelopes. For thin shells
the difference increases but always remains small compared to
the scatter of the observational points. This is illustrated in Fig.~2 where 
a model having
the inner to outer radius ratio equal to 0.95 is displayed (dotted curve).

\subsection{$r^{-2}$ density distribution}

In theoretical modelling of radio emission from circumstellar envelopes
it is sometimes adopted that the density of the emitting matter varies as
$r^{-2}$ (or $r^{-\alpha}$ in more general), where $r$ is the distance 
from the central star (e.g. Panagia
\& Felli 1975, Taylor et~al. 1987). This sort of density distribution
results from a simple steady-state wind model. We have also attempted to
compare this kind of model to our observational data.

The nebular matter is assumed to be distributed spherically symmetrically.
Starting from $R_{\rm in}$ the nebular density varies as $(r/R_{\rm in})^{-2}$
which implies that the absorption coefficient varies as 
$(r/R_{\rm in})^{-4}$. As in the previous subsection the flux has been
calculated from numerical integration of the radiation transfer equation.

The radio brightness temperature for a given model nebula has been derived
in a way similar to that usually applied for observed objects.
The surface brightness distribution across the nebular disc has been 
calculated. Then the nebular radius has been defined as a
distance from the centre of the nebular disc to the place 
where the surface brightness
drops to 10\% of its maximum value. With this radius and the flux (both done
for 5~GHz) the radio brightness temperature has been derived from
Eq.(\ref{tb1}).

The results are shown as a dashed curve in Fig.~2. As expected from
analytical considerations (e.g. Panagia \& Felli 1975) for large optical
thicknesses ($T_{\rm b}/T_{\rm e} \simeq 1$) $F_{\rm 5GHz}/F_{\rm 1.4GHz}$
approches a limit of 2.12. This is much smaller than the limiting values in
the previous models. The reason is that for optically thick $r^{-2}$
distributions the radius of the emitting nebular disc increases with the
decreasing frequency ($\sim \nu ^{-0.7}$ -- see e.g. Panagia \& Felli
1975). Thus although the surface brightness of the nebular disc is close to
that of the black body ($\sim \nu ^2$) the flux is less steep 
($\sim \nu ^{0.6}$). 

As can be seen from Fig.~2 the $r^{-2}$ model (dashed curve) does not fit
the observed positions. 
It makes the discrepancy between theory and observations even
larger when compared to the previous models. One may argue that it is more
realistic to consider $r^{-2}$ density distributions with an outer cutoff
due to the ionization front, as in Taylor et~al. (1987). Then the results
depend on the ratio of inner to outer radii, denoted by $\eta$ in Taylor et~al.
However, for $\eta \ga 0.3$ the results are very close to the 
dotted (uniform shell model) and full (basic
model) curves in Fig.~2. 
Only for $\eta < 0.3$ the $r^{-2}$ density distribution
deviates noticeably from the basic model but towards the dashed curve
(infinite $r^{-2}$).

Thus, contrary to Taylor et~al. (1987), we conclude that 
the $r^{-2}$ density distribution is not typical for
the planetary nebulae. 
This conclusion is consistent with what we can learn from observed
images and hydrodynamical modelling of planetary nebulae.
Images in optical emission lines (e.g. Balick 1987, Manchado et~al. 1996,
G\'orny et~al. 1999)
clearly show that they cannot be reconciled
with the $r^{-2}$ density distribution in the great majority of the
planetary nebulae. Hydrodynamical simulations of formation and evolution of
planetary nebulae (e.g. Marten \& Sch\"onberner 1991, Corradi et~al. 2000)
predict the density structure 
significantly different from the $r^{-2}$ law, as well.

\subsection{Two-component model}

It is well known that most of the planetary nebulae do not show spherical
symmetry. Often brighter and denser regions 
in form of bipolar structures, blobs
or condensations are embedded in fainter and thinner material. In order to
investigate this situation 
we have constructed a two-component model which is a sort of
modification of our basic model in Sect.~3.1. 

We adopt that the image of the nebula, while seen by an observer, 
consists of denser and more opaque regions
characterized by an optical thickness, $\tau_\nu$, and thinner regions having
the optical thickness of $\epsilon \tau_\nu$. The thicker regions fill up a
solid angle of $\xi \Omega$, where $\Omega$ is the solid angle of the
nebula as a whole. Correspondingly the thinner regions fill up 
$(1 - \xi)\Omega$. Both kinds of regions have the same electron
temperature, $T_{\rm e}$.

Within this model the observed flux, $F_\nu$, is given by
\begin{equation}
\label{fnu3}
F_\nu = {{{2\nu^2}kT_{\rm e}}\over {c^2}}((1-e^{-\tau_\nu})\xi\Omega
+(1-e^{-\epsilon\tau_\nu})(1-\xi)\Omega)
\end{equation}
while the ratio of $T_{\rm b}/T_{\rm e}$ can be obtained from
\begin{equation}
\label{tb3}
T_{\rm b}/T_{\rm e} = 1 - \xi e^{-\tau_\nu} - (1 - \xi)
e^{-\epsilon\tau_\nu}
\end{equation}

\begin {figure}
 \resizebox{\hsize}{!}{\includegraphics{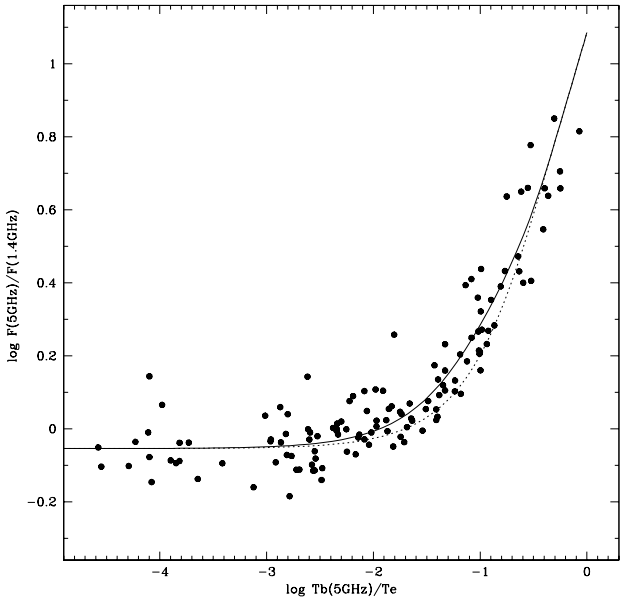}}
 \caption{Comparison of the basic model (dotted curve) and the
two-component model with $\xi=0.27$ and $\epsilon=0.19$ (full curve) 
with the observed
nebulae having more reliable determinations of the diameters (+ signs
in Table~1).}
\end{figure}

Depending on the adopted values of $\xi$ and $\epsilon$
the two-component models deviate more or less from the basic model
but always towards the upper-left in Fig.~2. 
Thus they can significantly improve the
agreement between the observations and the modelling. However, in order
to reach the objects lying well to the
upper-left from the basic model in Fig.~2 $\xi \la 0.1$ and $\epsilon \la
0.01$ have to be adopted. Perhaps it would be too simplistic to interpret
these objects as
consisting of a few small and dense blobs embedded in
an extended thin nebula. Although this possibility cannot be excluded the
extreme positions of these objects can also be due to inaccuracies in
the observational parameters. For instance, an overestimate of
the nebular diameter would result in a underestimate of $T_{\rm b}$.
Although an overestimate of the nebular diameter by a factor of a few
(required to explain the most extreme positions in Fig.~2) is rather
excluded for large, well resolved nebulae it may be the case for objects
comparable to or smaller than the observational resolution.

Therefore for a quantitative comparison between the observations and
the two-component models we have selected a subsample of objects for which
the diameters ("radio" and/or "optical") have been assigned with + in Table~1.
We expect (as explained in Sect.~2) that these objects have
relatively accurate values of the diameters. 
The resultant subsample (129 objects) is plotted in Fig.~3. 

\begin {figure}
 \resizebox{\hsize}{!}{\includegraphics{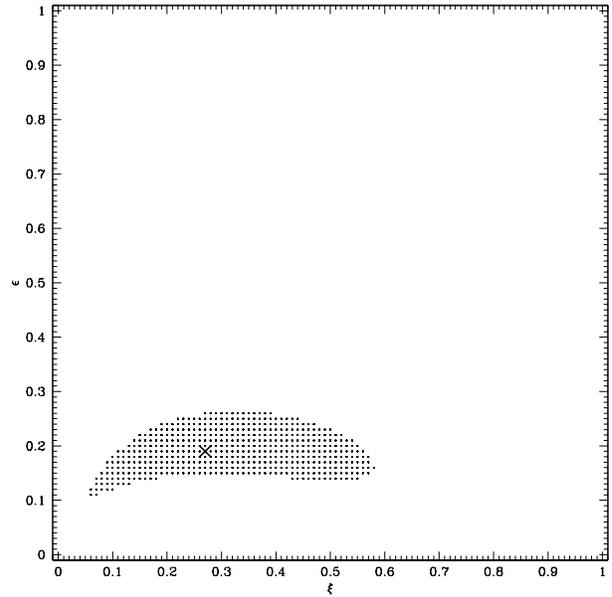}}
 \caption{The 90\% confidence level of the $\chi^2$ test of the
two-component model. The cross shows the best fit.}
\end{figure}

As can easily be seen the observed positions plotted in Fig.~3 are much
better reproduced by the basic model (dotted curve) than those 
displayed in Fig.~2. However, the fit is not perfect. For intermediate
values of log~$T_{\rm b}/T_{\rm e}$ the observed positions still tend to
be shifted to the upper-left from the dashed curve. A better fit to the
observations can be obtained from a two-component model. The full curve
shows a model with $\xi = 0.27$ and $\epsilon = 0.19$ which gives the best
fit to the observations according to the $\chi^2$ test. The $\chi^2$
calculations have been done adopting that the typical error of the observed
flux ratio is 0.075 in dex. This is the observed dispersion of the data 
for optically thin nebulae, i.e. having log~$T_{\rm b}/T_{\rm e} < -2.5$ 
in Fig.~3 (39 objects, the mean value of the flux ratio is $-0.049$ in dex).
The $\chi^2$ minimum
is, however, rather shallow. This can be seen from Fig.~4 which shows 
the 90\% confidence
level of the $\chi^2$ test in the $\xi - \epsilon$ plane. Thus, roughly
speaking, a reasonable fit of the model to the data in Fig.~3 is obtained
for $0.1 \la \xi \la 0.6$ and $0.15 \la \epsilon \la 0.25$.

It may seem, perhaps at first sight, that the two-component model (full
curve) makes only a minor improvement of the fit in Fig.~3 as compared to 
the basic model (dotted curve). That this is not the case one can see from
Fig.~4 which shows that the basic model ($\xi = 1.0$ and/or $\epsilon =
1.0$) is significantly apart from the 90\% confidence level of the
two-component model. The $\chi^2$ distance of the basic model from the best
two-component model ($\xi = 0.27$ and $\epsilon = 0.19$) is 39.2. This means
that the ratio of the confidence level for the basic model
to the confidence level for the best two-component model is much less than
0.01\%. Thus the two-component model presents a significant improvement over
the basic model.

\begin {figure}
 \resizebox{\hsize}{!}{\includegraphics{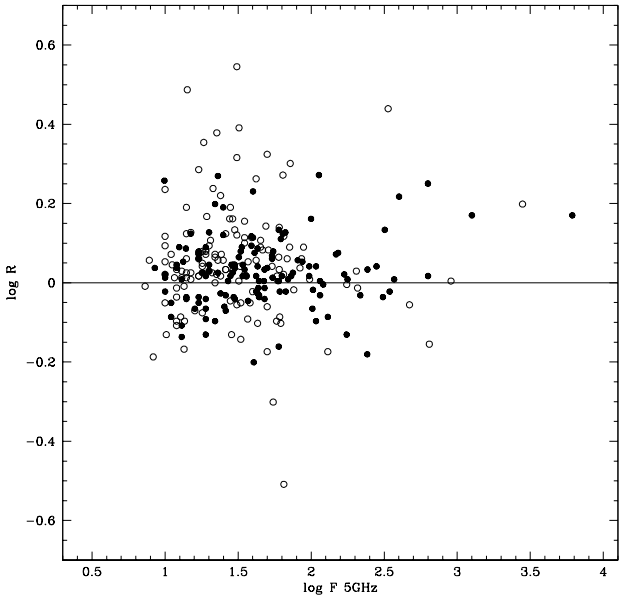}}
 \caption{$R$ versus the observed flux at 5~GHz.
  Full symbols: objects denoted with a +
  sign in column (6) of Table~1. Open symbols: other objects.}
\end{figure}

\section{Uncertainities in the observational data}

It is obvious that the observational data we are using in the present study
are not free of uncertainties. As we have concluded from the data for optically
thin nebulae in Sect.~3.1 a typical accuracy of the flux ratio in our sample
is 0.10 (in dex). However, the accuracy need not to be the same for the
whole sample. For instance, it can be expected that the fluxes are more 
uncertain for faint objects, while the diameters are less precise for small
nebulae. In this section we discuss possible sources of uncertainties in our
observational data.

We consider the two-component model with $\xi = 0.27$ and $\epsilon = 0.19$,
shown by the full curve in Fig.~3, as a "reference" model for our
sample. As found in Sect.~3.4 it reproduces satisfactorily 
the subsample of more accurately measured objects. Using the observed value
of $T_{\rm b}/T_{\rm e}$ for a given object we can predict the 
$F_{\rm 5GHz}/F_{\rm 1.4GHz}$ ratio from the reference model. Then we define 
$R$ as a ratio of the observed $F_{\rm 5GHz}/F_{\rm 1.4GHz}$ to the predicted 
$F_{\rm 5GHz}/F_{\rm 1.4GHz}$. Thus the value of $R$ is a measure of departure of
the data for a given object from the reference model. 
In the following we analyse whether this departure correlates with the
observed fluxes and diameters.

In Fig.~5 we have plotted $R$ versus the observed flux at 5~GHz. Full
symbols denote objects with more reliable measurements of the nebular
diameter (+ signs in column (6) or/and (8) of Table~1). In the 
whole range of log~$F_{\rm 5GHz}$ the mean value of log~$R$ for the
entire sample (264 objects) is within $0.03-0.05$ 
while the standard deviation is 0.12. The full symbols (129 objects) 
show a somewhat lower deviation, i.e. 0.09.

\begin {figure}
 \resizebox{\hsize}{!}{\includegraphics{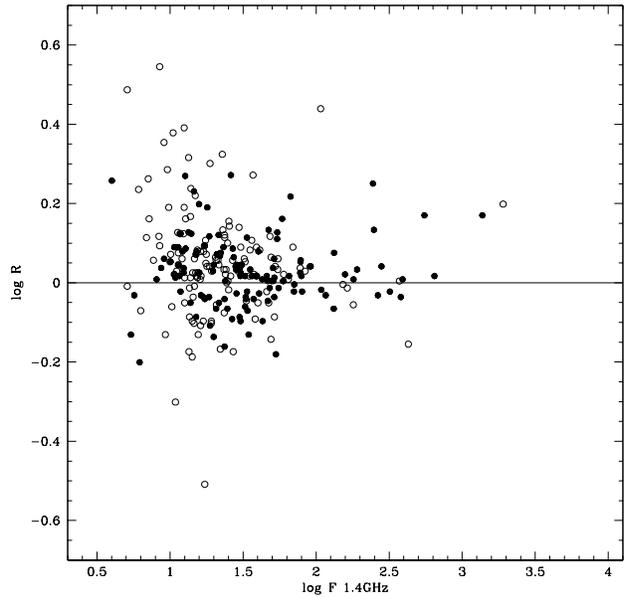}}
 \caption{$R$ versus the observed flux at 1.4~GHz.
  Full symbols: objects denoted with a +
  sign in column (6) of Table~1. Open symbols: other objects.}
\end{figure}

The same but for the observed flux at 
1.4~GHz is given in Fig.~6.  While the mean value of log~$R$ is here also 
within $0.03-0.05$ the standard deviation is higher for fainter objects. For 
log~$F_{\rm 1.4GHz} > 1.3$ in the whole sample (149 objects) the
standard deviation is 0.09 while is it equal to 0.15 for log~$F_{\rm 1.4GHz} < 1.3$ 
(115 objects). When only the full symbols are considered the standard
deviation is, respectively, 0.08 (88 objects) and 0.10 (41 objects).

The conclusion that can be drawn from Figs.~5 and 6 and the values of the
standard deviation is that perhaps certain measurements of $F_{\rm 1.4GHz}
\la 20$~mJy might be less accurate. However, since the large deviation 
for log~$F_{\rm 1.4GHz} < 1.3$ is primarily due to
the open points uncertainities in the diameters can also be involved.

A plot of $R$ against the adopted nebular diameter (column (10)
in Table~1) indicates that the standard deviation for nebulae
smaller than $10\arcsec$ is higher, especially if the objects without a +
sign in Table~1 are considered. Since our adopted diameters have been
derived from a compilation of different observational data and different 
methods in the following we will discuss separately the radio diameters and
the optical diameters.

\begin {figure}
 \resizebox{\hsize}{!}{\includegraphics{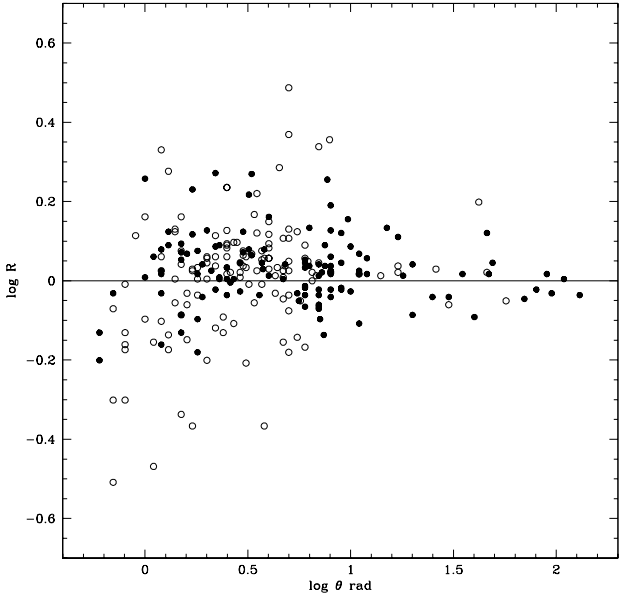}}
 \caption{$R$ versus the radio nebular diameter. 
  Full symbols: objects denoted with a +
  sign in column (6) of Table~1. Open symbols: other objects.}
\end{figure}

In Fig.~7 we have plotted the value of $R$ against the radio diameter. 
Note that also the radio (and not adopted) diameter has been used to derive
the value of $R$.
Full symbols correspond to the data having a + sign in 
column (6) of Table~1. As mentioned in Sect.~2 these values primarily come
from the 10\% radio contours.
Other objects, with measurements primarily resulting from a gaussian
deconvolution, are denoted with open cirles.
As can be seen the full circles in Fig.~7 give a fairly consistent result.
In the whole range of log~$\Theta$ the mean
value of log~$R$ is close to 0.02. For log~$\Theta > 1.0$ (26 objects) 
the standard deviation is
0.06 comparing to the value of 0.09 obtained for log~$\Theta < 1.0$ (95
objects). The other
radio diameters (open symbols in Fig.~7) give more dispersed results.
For log~$\Theta > 1.0$ (9 objects) the mean log~$R$ is 0.03 with 
a standard deviation of 0.07. However,
for log~$\Theta < 1.0$ (122 objects) the corresponding values
are 0.01 (mean) and 0.16 (standard deviation). Thus we can conclude that the
radio diameters for small nebulae, especially if they are derived from 
the gaussian fitting, are less certain. However, they do not show any
significant systematic effect in the results.

\begin {figure}
 \resizebox{\hsize}{!}{\includegraphics{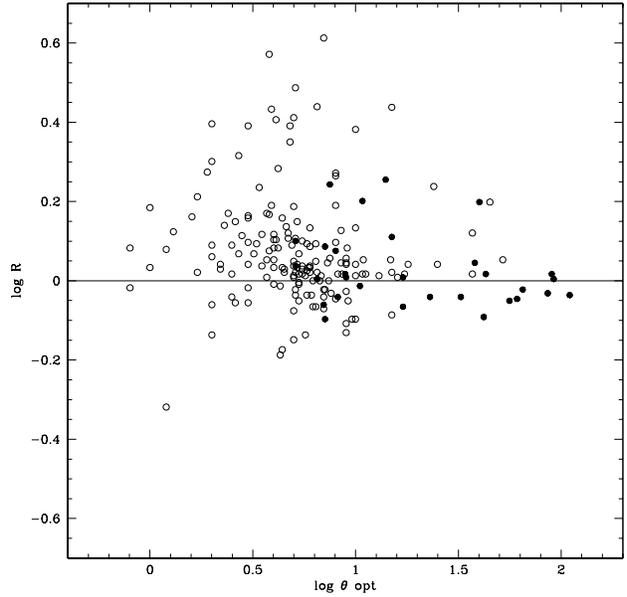}}
 \caption{$R$ versus the optical nebular diameter.
  Full symbols: objects denoted with a + sign in column (8) of Table~1. Open
  symbols: other objects. Objects with upper limits or
  denoted with ST. in column (8) of Table~1 are not included.}
\end{figure}

Fig.~8 shows the same as Fig.~7 but using the
optical diameters. Objects with upper limits or denoted with ST. in column
(8) of Table~1 are excluded from the diagram. Similarily as in the previous 
figure the distribution of the full symbols in Fig.~8 is independent of the
diameter although the statistics is here lower (30 objects only) and concerns
relatively large nebulae. Nevertheless, for both, log~$\Theta > 1.0$ and
log~$\Theta < 1.0$, the mean value of log~$R$ is $0.02-0.03$ and the standard
deviation is $0.09-0.10$. However the open symbols which primarily concern
smaller nebulae show a larger spread and
tend to be systematically above the line log~$R = 1.0$.
For log~$\Theta > 1.0$ (23 objects) we have obtained the mean value of log~$R$ 
equal to 0.11 and the standard deviations of 0.21. For log~$\Theta < 1.0$ 
(145 objects) the result is $0.08 \pm 0.14$. The optical measurements for
the objects denoted with the open symbols have mostly been derived from
photographic plates. We conclude that they are often uncertain and tend to
overestimate the diameters.

Summarizing the present section, it seems that the accuracy of the 
flux measurements is more or less
uniform in our sample. Perhaps only for some objects the flux at 1.4~GHz
measured to be below 20~mJy is more uncertain. The accuracy of the diameter
measurements is however significantly worse for nebulae smaller than
$10\arcsec$. This particularly concerns determinations from low angular
resolution observations, i.e. results from
photographic plates and gaussian fitting to radio profiles. 

\section{Self absorption effects and derivation of interstellar extinction}

The value of interstellar extinction to the planetary nebulae is often
derived from an optical to radio flux ratio. This is usually done assuming
that the nebulae are optically thin. It is quite obvious that this is
not the case at 1.4~GHz for many objects in our sample. The fact that 
$F_{\rm 5GHz}/F_{\rm 1.4GHz}$ is often measured to be well above the optically
thin limit is primarilly due to self absorption at 1.4~GHz. 
At 5~GHz self absorption is always less important but it does not mean that 
it is always negligible.

\begin {figure}
 \resizebox{\hsize}{!}{\includegraphics{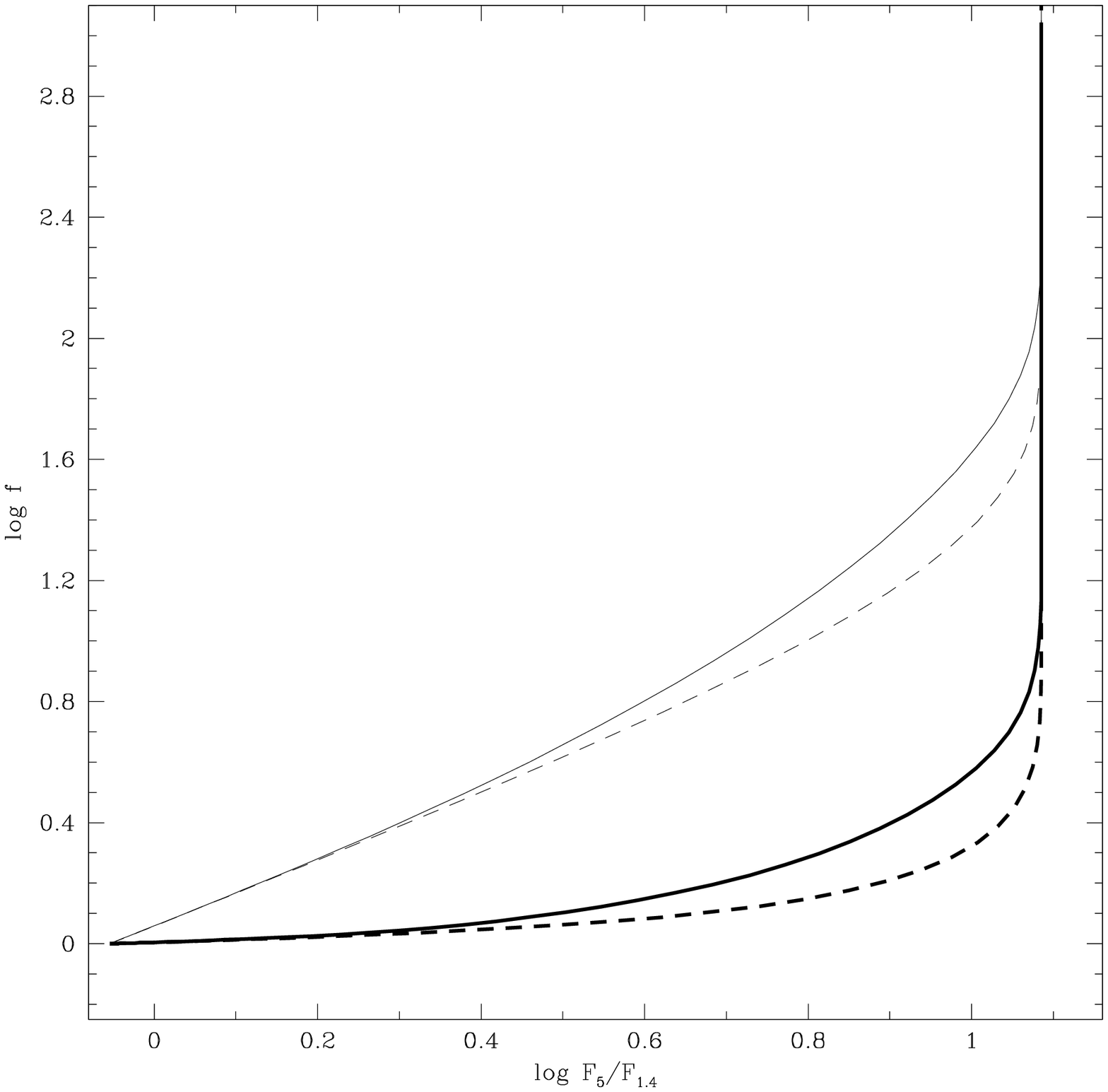}}
 \caption{The self absorption correction factor $f$ against 
$F_{\rm 5GHz}/F_{\rm 1.4GHz}$. Thin curves: $f$ at 1.4~GHz, thick curves: 
$f$ at 5~GHz. Dashed curves: basic model, full curves: reference model 
(two-component model with $\xi = 0.27$ and $\epsilon = 0.19$).}
\end{figure}

\begin {figure}
 \resizebox{\hsize}{!}{\includegraphics{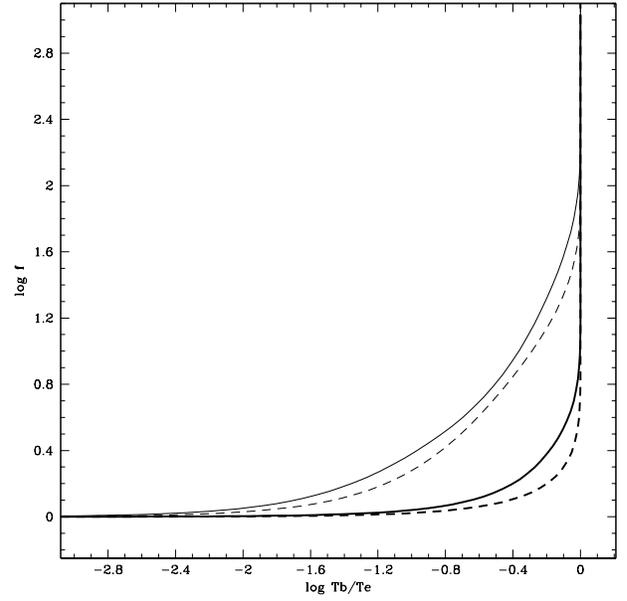}}
 \caption{The self absorption correction factor $f$ against 
$T_{\rm b}/T_{\rm e}$. 
Thin curves: $f$ at 1.4~GHz, thick curves: $f$ at 5~GHz.
Dashed curves: basic model, full curves: reference model (two-component
model with $\xi = 0.27$ and $\epsilon = 0.19$).}
\end{figure}

Let us define $f$ as
a ratio of the flux obtained adopting negligible optical
thickness (i.e. from simple integration of the emission coefficient over the
nebula) to that derived with self absorption taken into account. If 
$f$ can be estimated than the observed radio flux has to be 
multiplied by this factor before deriving interstellar extinction. In
principle the value of $f$ can be derived either from 
$F_{\rm 5GHz}/F_{\rm 1.4GHz}$ or from $T_{\rm b}/T_{\rm e}$ as both parameters
measure optical thickness effects.

In Fig.~9 we have plotted $f$ at 1.4~GHz (thin curves) and 5~GHz (thick curves) 
against $F_{\rm 5GHz}/F_{\rm 1.4GHz}$ calculated from the basic model 
(dotted curves) and the reference model (full curves).
The same but displayed against $T_{\rm b}/T_{\rm e}$ is shown in
Fig.~10. (Note that $T_{\rm b}$ in Fig.~10, as throughout in this paper, 
refers to 5~GHz.)

As can be seen from Figs.~9 and 10 the relations between $f$ and the
observational quantities are model dependent.
Generally speaking, for a given value of the flux ratio or 
$T_{\rm b}/T_{\rm e}$ the
two-component model predicts a larger $f$ than the basic model.
For the two-component model the relations in Figs.~9 and 10 depends 
on the model parameters, i.e. $\xi$ and $\epsilon$.

We have estimated $f$ for the objects in our sample.
We have adopted the
reference model ($\xi = 0.27$ and $\epsilon = 0.19$).
$f$ has been estimated from 
the observed $F_{\rm 5GHz}/F_{\rm 1.4GHz}$ and
$T_{\rm b}/T_{\rm e}$, and the mean value from the two has been taken.
Fig.~11 plots the flux at 5~GHz corrected for self absorption, i.e.
observed multiplied by $f$, against the observed flux. The same but for
1.4~GHz is shown in Fig.~12.

\begin {figure}
 \resizebox{\hsize}{!}{\includegraphics{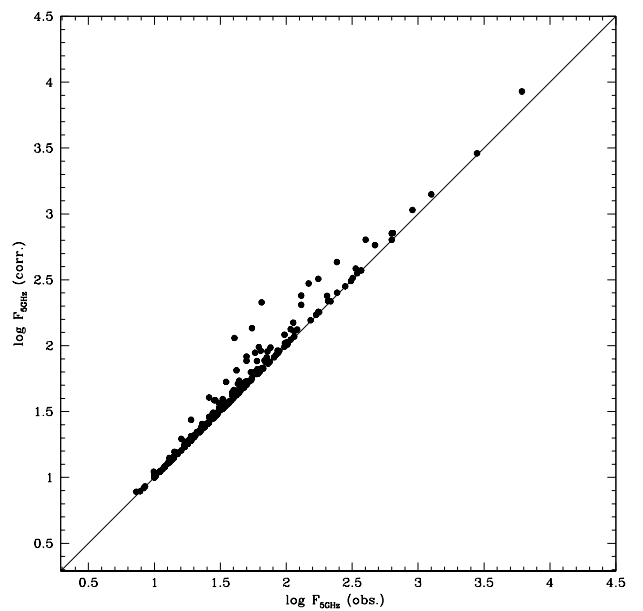}}
 \caption{The flux at 5~GHz corrected for self absorption against the
observed flux. Full line: the 1:1 relation.}
\end{figure}

As can be seen from Fig.~11, for most of the objects the self absorption at
5~GHz is not important. In a few cases, however, it can be as large as
factor 2. As expected, the self absorption at 1.4~GHz is much more
important. As can be seen from Fig.~12, in a number of cases it is of the
order of 10.

Estimates of the self absorption factor are usually done in the literature
from the observed $T_{\rm b}$ and formulae which are equivalent to our basic
model (e.g. Condon \& Kaplan 1998). As can be seen from Fig.~10 this leads
to smaller correction factors than those from our procedure based on the
reference model. We have repeated our estimates as described above
but using the basic model. The resultant correction factors, 
$f_{\rm basic}$, are indeed smaller than the previously derived $f$ values.
The relation between the two estimates can be very well approximated by 
${\rm log} f_{\rm basic} = \alpha \,{\rm log} f$. For the flux at 5~GHz
$\alpha \simeq 0.61$  whereas at 1.4~GHz $\alpha \simeq 0.87$. Thus at
1.4~GHz where $f$ can be as large as 10 the basic model
predicts the correction factor about 30\% lower than the reference model.

\begin {figure}
 \resizebox{\hsize}{!}{\includegraphics{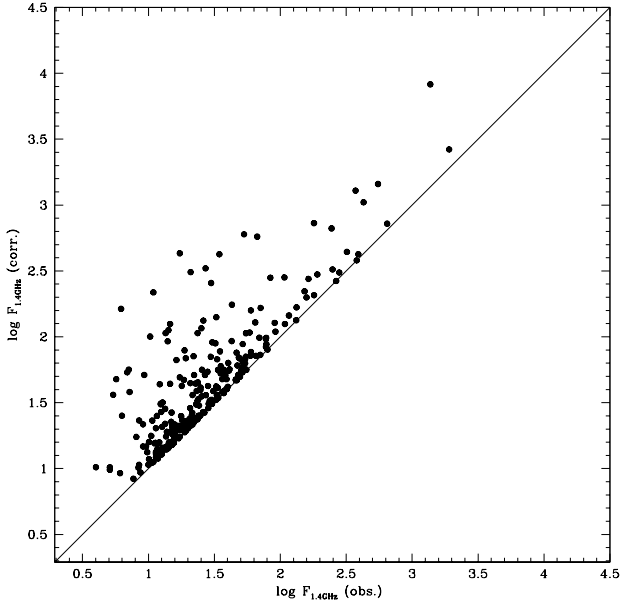}}
 \caption{The flux at 1.4~GHz corrected for self absorption against the
observed flux. Full line: the 1:1 relation.}
\end{figure}

As explained above our procedure for deriving $f$ relies on the flux ratio and
the brightness temperature at 5~GHz.
In principle, one can also use the brightness temperature at 1.4~GHz,
as done, for instance, in Condon \& Kaplan (1998) while correcting 
the fluxes at 1.4~GHz. 
The advantage of our approach is that it allows to estimate the self
absorption even for quite opaque nebulae at 1.4 GHz. As can be seen from
Figs.~9 and 10, from both, the flux ratio and $T_{\rm b}$ at 5~GHz,
the correction factor at 1.4 GHz (thin curves) can be reliably estimated up 
to values of 20-30. The relation
between $f$ at 1.4~GHz and $T_{\rm b}$ at 1.4~GHz is described by the thick
curves in Fig.~10 (the same as for both quantities at 5~GHz).
Clearly it becomes useless for the self absorption greater 
than 2-3.

Stasi\'nska et~al. (1992) have found that there is a systematic difference
between the extinction derived from the Balmer decrement, $C_{\rm opt}$, 
and from the ratio of 5~GHz to H$\beta$ fluxes, $C_{\rm rad}$. We have 
repeated their analysis for objects in our sample adopting our correcting 
precedure for self absorption at 5~GHz. 
The logarithmic extinction at H$\beta$, $C_{\rm rad}$, has been
calculated in a standard way (e.g. Pottasch 1984) using the observed
H$\beta$ fluxes from Acker et~al. (1992). The HeII$\lambda 4686$\AA~line 
intensities from Tylenda et~al. (1994) have been used in order to
estimate the electron temperature and the ionization degree of helium.
The observed values of H$\alpha$/H$\beta$ have been taken from
Tylenda et~al. (1994). We have excluded objects for which the accuracy of
the H$\beta$ flux (Acker et~al. 1992) is worse than 0.1 in dex. 

\begin {figure}
 \resizebox{\hsize}{!}{\includegraphics{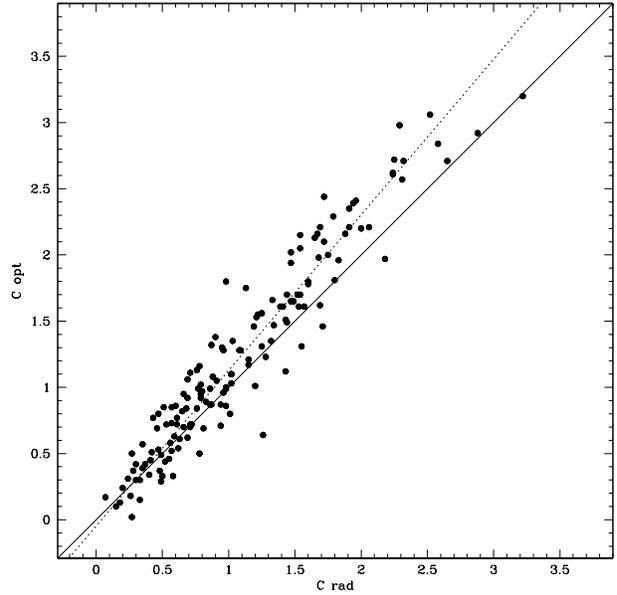}}
 \caption{$C_{\rm opt}$ derived from the Balmer decrement versus 
$C_{\rm rad}$ from the 5~GHz to H$\beta$ flux ratio. Solid line: the
1:1 relation, dashed line: the orthogonal regression line.}
\end{figure}

The results
are displayed in Fig.~13 which can be compared to Fig.~5 in Stasi\'nska
et~al. The orthogonal regression line for the points in Fig.~13 is
(neglecting, as in Stasi\'nska et~al., the point corresponding to IC~4997: 
$C_{\rm rad} \simeq 0, C_{\rm opt} \simeq 1$)
\[
C_{\rm opt} = 1.175\, C_{\rm rad} - 0.046.
\]

Thus we have obtained the effect very close to that in Stasi\'nska et~al.
This is as expected since the self absorption at 5~GHz is negligible 
for most of the objects. Nevertheless 
our relation between $C_{\rm opt}$ and $C_{\rm rad}$
is a bit less steep. The least squares straight line passing through
the origin in Fig.~13 has a slope of 1.133 compared to 1.17 derived in
Stasi\'nska et~al.

\section{Shklovsky distances to the planetary nebulae in the Galactic bulge}

One of the most widely used (and most often critisized) methods for 
determining distances to the planetary nebulae is the Shklovsky method. The
basic assumption in this method is that the nebulae have the same mass,
usually adopted to be $0.2\, M_{\odot}$. This is probably the most important
source of uncertainties in the derived distances as the nebulae are expected
to have an important spread in their masses.

A test (or a calibration) of a method for determining distances can be done 
if it can be applied to a sample of objects with known distances. 
This is the case of the planetary nebulae in the Galactic bulge. 
This sort of test of the Shklovsky method has been done in Stasi\'nska
et~al. (1991) and in Pottasch \& Zijlstra (1992). The two groups have drawn
rather opposite conclusions. The Shklovsky distances derived in 
Stasi\'nska et~al. are compatible (for most of the objects within factor 2) 
with the known distance to the Galactic centre. Pottasch \& Zijlstra
conclude that the Shklovsky method is wrong as it systematically
overestimates the distances.

We have calculated the Shklovsky distances for a sample of the Galactic
bulge objects selected from Table~1. The selection has been done in a
standard way, i.e. nebulae within $10\degr$ from the Galactic centre having
the diameter smaller than $20\arcsec$ and the flux at 5\,GHz lower than 100~mJy.
This gave as a sample of 80 objects which is smaller than in the two
previous studies. The reason is that our Table~1 includes only objects with
reliable flux measurements at 5 and 1.4\,GHz.

We have used the same formula as in Stasi\'nska et~al. (1991). 
The nebulae are assumed to have a mass of $0.2\, M_{\odot}$
and a filling factor of 0.5. The observed H$\beta$ fluxes (Acker et~al.
1992) have been corrected for the interstellar extinction, $C_{rad}$, using
the 5\,GHz fluxes. Note that the radio fluxes have been corrected for
self absorption and that the effects of the electron temperature and the He
ionization degree have been taken into account, as explained in Sect.~5.
We have used the "adopted" diameters from column (10) of Table~1.
Note that the result would be practically the same
if we took the "radio" values from Table~1 (column 6). This is not
surprising since, as noted in Sect.~2, for small nebulae we have
preferencially adopted the values from radio measurements.

\begin {figure}
 \resizebox{\hsize}{!}{\includegraphics{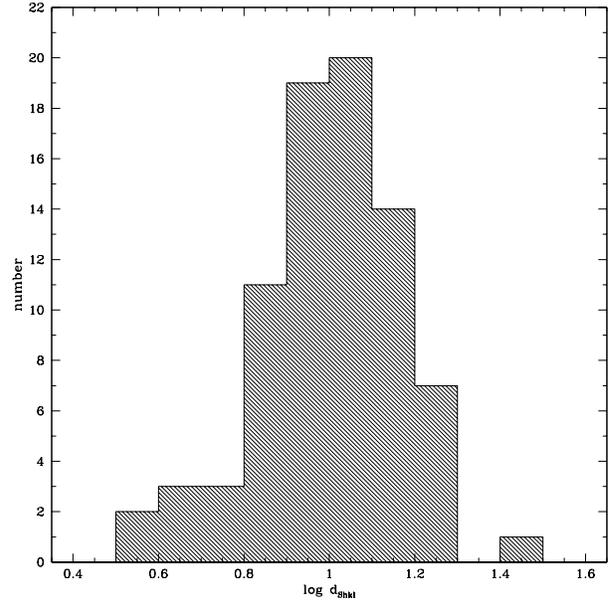}}
 \caption{Shklovsky distances to the Galactic bulge objects.}
\end{figure}

Fig.~14 shows a histogram of the obtained Shklovsky distances, 
$d_{\rm Shkl}$. The mean value of 
log$\,d_{\rm Shkl}$ is 0.997 and the standard deviation is 0.167. 
The derived mean value (9.9~kpc) is somewhat larger than the current
estimates of the
distance to the Galactic centre which varies from 7 to 9~kpc depending on
the method used (Reid 1999). However, we do not conclude that the
Shklovsky method is wrong. Fig.~14 can be used for calibrationg the method
and for estimating its accuracy. 

The mean value of the Shklovsky distances would be equal to 8.5~kpc (the
IAU-approved value for the distance to the Galactic centre)
if the nebular mass adopted in the Shklovsky 
formula were $0.14\, M_{\odot}$. In other words, if the Galactic bulge
planetary nebulae are used to calibrate the Shklovsky method than the
nebular mass of $0.14\, M_{\odot}$ should be adopted in the distance formula. 
An open question is, however, to what extend the planetary nebulae in the
Galactic bulge can be regarded as representative for the whole Galactic
population. 

The standard deviation of the log$\,d_{\rm Shkl}$ distribution can be
considered as an average error of the Shklovsky method. The distribution in
Fig.~14 would thus imply that the Shklovsky distances are typically
uncertain by factor 1.5. In the worst individual cases the error can reach a
factor of 3. The internal uncertainty of the Shklovsky
method (resulting primarily from the assumption of a constant nebular mass) is
however smaller. In the case of the Galactic bulge nebulae the observational
errors can be important, especially in the diameters to which the Shklovsky
distances are most sensitive. The observed standard deviation in the
Shklovsky distances could be entirely accounted for by a standard deviation
in the diamaters equal to 0.3 in dex. A typical uncertainty of factor 2 in
the diameters is probably too pessymistic. Nevertheless, the uncertainty is
expected to be important as the Galactic bulge
nebulae are in their majority smaller than $10\arcsec$ and their
diameters primarily come from the radio gaussian fitting (cf.
discussion of Fig.~7 in Sect.~4).

\section{Summary and conclusions}

We have analysed the radio fluxes for 264 planetary nebulae. These are
objects for which reliable measurements of fluxes at 1.4 and 5~GHz from VLA
are available. Also nebular dimensions are known for them. 

We have shown
that for many of the investigated nebulae the optical thickness
is important, especially at 1.4~GHz. We have attempted different nebular
models in order to account for the observations. A simple model specified
only by a single optical thickness is not satisfactory. The same has been
concluded for spherically symmetric, constant density shells.
A spherically symmetric models with an $r^{-2}$ density
distribution has also been ruled out. A reasonable representation of 
the observations
has been obtained from a two-component model in which the nebular image
consists of regions having two different values of optical thickness. 

We have analysed uncertainties in the observational data. The accuracy of
the flux measurements is more or less uniform in our sample. Perhaps for
some objects the flux at 1.4~GHz, if it is below
20~mJy, can be more uncertain. 
We have found that the uncertainity in the nebular diameters
siginificantly increases for nebulae smaller that $10\arcsec$. 
This concerns the measurements from photographic plates
and from gaussian fitting to the radio profile. 

While determining the interstellar
extinction from an optical to radio flux ratio caution should be paid to
optical thickness effects in the radio.
We have developed a method for estimating the value of self absorption.
It relies on the observed radio flux ratio (5~GHz to 1.4~GHz) and the
observed brightness temperature at 5~GHz. The results, however, depend on
the adopted nebula model.
At 1.4~GHz self absorption of the flux is usually important and can exceed
a factor of 10. At 5~GHz
self absorption is negligible for most of the objects although in some cases
it can reach a factor of 2.
Similarily as in Stasi\'nska et~al.
(1992) we have obtained a systematic
difference between the extinction derived from the Balmer decrement and that
from the optical to radio flux ratio. Our relation is however slightly less
steep due to (usually small) corrections for self absorption of the 5~GHz
fluxes in our study.

We have attempted to calibrate the Shklovsky method using the planetary
nebulae in the Galactic bulge. The mean value of the Shklovsky distances 
is close to the distance to the Galactic centre if the nebular mass is
adopted to be
$0.14\, M_{\odot}$. This is smaller than a canonic value of
$0.2\, M_{\odot}$ usually adopted in the Shklovsky method. Unfortunately the
question whether the Galactic bulge nebulae are typical for the whole
Galactic population is still open. From the distribution of the Shklovsky
distances for the Galactic bulge nebulae we have derived that
the statistical uncertainity of the Shklovsky distances
should be smaller than factor 1.5. 

A study like that done in this paper could be much more accurate and
conclusive if an extensive survey of the planetary nebulae, similar to those
existing at 5 and 1.4~GHz, were available at a higher frequency. 
The fluxes at three frequencies
would allow to better test different model nebulae and to study more
deeply the optical thickness effects. The flux at the higher frequency could
be used with great confidence even for small and compact nebulae to derive
interstellar extinction. In fact
there is a survey at 14.7~GHz by Milne \& Aller (1982) but this
has been done with the Parkes telescope. 
In view of our experiences (this paper as
well as Stasi\'nska et~al., 1992) high frequency
measurements with the VLA would be desirable. However, the
synthesized beam of the VLA, even in the most compact configuration,
is very small at high frequencies. Thus, as discussed in Pottasch \&
Zijlstra (1994), the fluxes of
extended planetary nebulae from this instrument could be systematically
underestimated. Perhaps a survey with a
large single dish, e.g. the 100~m Green Bank Telescope, could be more
appropriate.

\begin{acknowledgements}
We thank the referee, J.J. Condon, whose comments resulted in a significant
improvement of the paper.
The research reported in this paper has partially been supported from the
grant 2.P03D.020.17 financed by the Polish State Committee for Scientific
Research.
\end{acknowledgements}

\begin{table*}
\centering
 \caption{Observational data.}
\begin{tabular}{p{1.8cm}p{2.0cm}r@{.}p{0.2cm}r@{.}p{0.5cm}p{1.0cm}r@{.}p{0.6cm}p{0.8cm}r@{.}p{0.5cm}p{1.1cm}r@{.}p{0.4cm}}
 \hline
  \multicolumn{1}{c}{PN G} & 
  \multicolumn{1}{l}{Main Name} &
  \multicolumn{2}{l}{F(1.4GHz)} & 
  \multicolumn{2}{l}{F(5GHz)} &
  \multicolumn{1}{l}{ref.} & 
  \multicolumn{2}{c}{diameter} &
  \multicolumn{1}{l}{ref.} & 
  \multicolumn{2}{c}{diameter} &
  \multicolumn{1}{l}{ref.} & 
  \multicolumn{2}{c}{diameter} \\
   \multicolumn{1}{c}{ } & 
   \multicolumn{1}{l}{ } & 
   \multicolumn{2}{c}{mJy} &
   \multicolumn{2}{c}{mJy} & 
   \multicolumn{1}{l}{ } &
   \multicolumn{2}{c}{(radio)} & 
   \multicolumn{1}{l}{ } &
   \multicolumn{2}{c}{(optical)} & 
   \multicolumn{1}{l}{ } &
   \multicolumn{2}{c}{(adopted)} \\
    \multicolumn{1}{c}{ } & 
    \multicolumn{1}{l}{ } & 
    \multicolumn{2}{l}{ } &
    \multicolumn{2}{l}{ } & 
    \multicolumn{1}{l}{ } & 
    \multicolumn{2}{c}{arcsec} &
    \multicolumn{1}{l}{ } & 
    \multicolumn{2}{c}{arcsec} & 
    \multicolumn{1}{l}{} & 
    \multicolumn{2}{c}{arcsec} \\
     \multicolumn{1}{c}{(1)} & 
     \multicolumn{1}{c}{(2)} & 
     \multicolumn{2}{c}{(3)} &
     \multicolumn{2}{c}{(4)} & 
     \multicolumn{1}{c}{(5)} & 
     \multicolumn{2}{c}{(6)} &
     \multicolumn{1}{c}{(7)} & 
     \multicolumn{2}{c}{(8)} & 
     \multicolumn{1}{c}{(9)} & 
     \multicolumn{2}{c}{(10)} \\
\hline
 000.1+04.3      & H 1-16   &  34&9 &  58&4 & 12      &  1&8  & 1    & $<$ 5&   & a       &  2& \\
 000.1+17.2      & PC 12    &  15&1 &  19&  & 1       &  1&8  & 1    &     2&5  & u       &  2&2 \\
 000.2\,--\,01.9 & M 2-19   &  10&5 &  22&6 & 2       &  5&   & 1    &    10&   & u       &  7&5 \\
 000.3+12.2      & IC 4634  & 116&4 & 115&  & 3, 4, 5 &  5&5 +& 3    &     8&6  & a       &  5&5 \\
 000.3\,--\,04.6 & M 2-28   &  12&6 &  10&  & 1       &   &   &      &     5&3  & a, b    &  5&3 \\
 000.7+03.2      & He 2-250 &  15&6 &  15&  & 6       &  5&   & 1    &     5&5  & a       &  5& \\
 000.7+04.7      & H 2-11   &  12&8 &  27&7 & 12      &  1&5  & 12   &     2&7  & c       &  1&5 \\
 000.9\,--\,04.8 & M 3-23   &  30&3 &  28&  & 1       & 12&0 +& 1    &    10&8  & a, b    & 11&5 \\
 001.0+01.9      & K 1- 4   &  21&5 &  25&  & 1       & 46&0 +& 1    &    37&   & a       & 42& \\
 001.2+02.1      & He 2-262 &  24&2 &  26&  & 1       &  3&1  & 1    &     4&   & d       &  3&1 \\
 001.2\,--\,03.0 & H 1-47   &   8&5 &  10&  & 6       &  2&5  & 1    & $<$ 5&   & a       &  2&5 \\
 001.3\,--\,01.2 & Bl M     &   9&6 &  17&  & 1       &  4&5  & 1    &     4&2  & a       &  4&5   \\
 001.4+05.3      & H 1-15   &  13&8 &  13&  & 1       &  4&3  & 1    &     5&3  & a       &  4&3   \\
 001.5\,--\,06.7 & SwSt 1   &  27&1 & 130&  & 3       &  1&3  & 3    & $<$ 5&   & a       &  1&3  \\
 001.7+05.7      & H 1-14   &  24&4 &  22&  & 1       &  6&5  & 1    &     6&7  & a       &  6&5   \\
 002.0\,--\,06.2 & M 2-33   &  12&3 &  12&  & 6       &  4&   & 1    &     5&2  & a, e    &  5&0   \\
 002.0\,--\,13.4 & IC 4776  &  37&0 &  64&1 & 4       &  1&6  & 4    &     8&   & f       &  8&    \\
 002.1\,--\,04.2 & H 1-54   &  13&4 &  31&  & 6       &  1&2  & 1    &     4&8  & g, b, e &  2&5   \\
 002.2\,--\,09.4 & Cn 1-5   &  51&7 &  44&  & 1       &  7&0 +& 1    & $<$ 5&   & a       &  7&    \\
 002.4\,--\,03.7 & M 1-38   &  14&9 &  24&  & 6       &  3&5  & 1    & $<$ 3&3  & a       &  3&5   \\
 002.6+04.2      & Th 3-27  &  14&7 &  13&5 & 5       &  3&2  & 5    &    10&   & i       &  3&2   \\
 002.6+08.1      & H 1-11   &  22&1 &  13&5 & 1, 5    &  6&   & 1, 5 &     4&4  & a       &  6&    \\
 002.6\,--\,03.4 & M 1-37   &  11&3 &  15&  & 6       &  2&5  & 1    &    ST&   & h       &  2&5   \\
 002.7\,--\,04.8 & M 1-42   &  28&6 &  28&5 & 2       &  9&0 +& 1    &     9&0  & a, b    &  9&    \\
 002.8+01.7      & H 2-20   &  13&8 &  19&3 & 13, 5   &  3&4  & 5    &     3&7  & a       &  3&5   \\
 003.1+02.9      & Hb 4     & 158&0 & 168&5 & 1, 3    &  7&25+& 1, 3 &     6&2  & a       &  7&25  \\
 003.1+03.4      & H 2-17   &   8&4 &  10&  & 1       &  4&   & 1    &     4&0  & u       &  4&    \\
 003.5\,--\,02.4 & IC 4673  &  54&1 &  62&  & 1       & 17&0 +& 1    &    15&0 +& f, a    & 16&    \\
 003.5\,--\,04.6 & NGC 6565 &  47&4 &  38&2 & 4       &  1&5  & 4    &     9&2  & a       &  9&2   \\
 003.6+03.1      & M 2-14   &  23&7 &  39&1 & 12      &  2&2  & 12   &    ST&   & h       &  2&2   \\
 003.7\,--\,04.6 & M 2-30   &  14&3 &  14&  & 1       &  3&5  & 1    &     9&0  & g, b    &  3&5   \\
 003.9\,--\,02.3 & M 1-35   &  54&7 &  60&5 & 2       &  4&4  & 1    &     4&4  & a, b    &  4&4   \\
 004.6+06.0      & H 1-24   &   5&1 &  14&2 & 1, 5    &  5&0  & 1, 5 &     5&1  & u       &  5&0   \\
 004.9+04.9      & M 1-25   &  40&3 &  55&  & 3, 4    &  3&2 +& 3    &     4&6  & a       &  3&2   \\
 005.2+05.6      & M 3-12   &  11&4 &  12&5 & 1       &  7&5 +& 1    &     6&0  & a       &  7&5   \\
 005.5+06.1      & M 3-11   &  11&8 &  10&  & 1       &  7&0 +& 1    &     7&1  & a       &  7&    \\
 005.8\,--\,06.1 & NGC 6620 &  17&5 &  20&5 & 2       &  5&   & 1    &     4&7  & a       &  5&    \\
 006.0\,--\,03.6 & M 2-31   &  41&2 &  51&  & 6       &  4&   & 1    &     5&1  & g, b, e &  4&    \\
 006.1+08.3      & M 1-20   &  33&2 &  48&  & 6, 3, 5 &  1&9 +& 3    & $<$ 7&   & i       &  1&9   \\
 006.2\,--\,03.7 & KFL 15   &  11&4 &  11&2 & 14      &   &   &      &     8&5  & b       &  8&5   \\
 006.3+04.4      & H 2-18   &  10&1 &  11&  & 6       &  3&   & 1    &     3&8  & a       &  3&5   \\
 006.4+02.0      & M 1-31   &  29&7 &  60&5 & 2       &  7&0  & 1    &     2&3  & u       &  2&3   \\
 006.8+02.3      & Th 4- 7  &   9&1 &  18&4 & 5       &  7&9  & 5    & $<$ 6&   & h       &  7&    \\
 006.8+04.1      & M 3-15   &  48&4 &  65&  & 6       &  5&   & 1    &     4&1  & a       &  4&5   \\
 007.0\,--\,06.8 & VY 2- 1  &  32&7 &  37&  & 1       &  3&7  & 1    & $<$ 7&   & a       &  3&7   \\
 007.2+01.8      & Hb 6     & 190&8 & 243&  & 1, 4    &  6&0 +& 1    &     5&8  & a       &  6&    \\
 007.5+07.4      & M 1-22   &   7&7 &   7&8 & 2       &  6&   & 1    &     9&   & a       &  6&    \\
 007.8\,--\,04.4 & H 1-65   &   6&1 &  10&  & 6       &  2&5  & 1    &     8&   & h       &  2&5   \\
 008.2+06.8      & He 2-260 &   8&1 &  13&  & 3       &  1&0 +& 3    &     1&2  & u       &  1&    \\
 008.2\,--\,04.8 & M 2-42   &   9&8 &  14&  & 1       &   &   &      &     3&9  & a       &  3&9   \\
 008.3\,--\,01.1 & M 1-40   & 163&6 & 208&  & 1       &  4&3  & 1    &     5&1  & a       &  4&5   \\
 008.3\,--\,07.3 & NGC 6644 &  64&6 &  97&5 & 6, 4    &  3&   & 1    &     2&5  & a, f    &  2&8   \\
 009.0+04.1      & Th 4- 5  &  20&7 &  16&  & 1       &  7&0 +& 1    &     6&4  & d       &  7&    \\
 009.4\,--\,09.8 & M 3-32   &  12&4 &  12&  & 1       &  7&5 +& 1    &     5&9  & a       &  7&5   \\
 009.6+14.8      & NGC 6309 & 132&1 & 102&  & 5       & 30&   & 5    &    17&0 +& a, z    & 17&    \\
 009.8\,--\,04.6 & H 1-67   &  12&6 &  11&6 & 1, 5, 2 &  6&8  & 5    &     5&7  & a       &  6&    \\
\hline
 \multicolumn{15}{r}{continued on next page}  
\end{tabular}
\end{table*} 

\begin{table*}
\centering
\begin{tabular}{p{1.8cm}p{2.0cm}r@{.}p{0.2cm}r@{.}p{0.5cm}p{1.0cm}r@{.}p{0.6cm}p{0.8cm}r@{.}p{0.5cm}p{1.1cm}r@{.}p{0.4cm}}
 \multicolumn{15}{l}{continued from previous page} \\ 
\hline
    \multicolumn{1}{c}{(1)} &
    \multicolumn{1}{c}{(2)} & 
    \multicolumn{2}{c}{(3)} &
    \multicolumn{2}{c}{(4)} & 
    \multicolumn{1}{c}{(5)} & 
    \multicolumn{2}{c}{(6)} &
    \multicolumn{1}{c}{(7)} & 
    \multicolumn{2}{c}{(8)} & 
    \multicolumn{1}{c}{(9)} & 
    \multicolumn{2}{c}{(10)} \\
\hline
 010.1+00.7      & NGC 6537 & 428&  &  644&  & 7       &   4&7  & 7    &    10&   & a       &   4&7   \\
 010.4+04.5      & M 2-17   &  10&7 &   10&  & 1       &   8&0 +& 1    &     6&5  & a       &   8&    \\
 010.7\,--\,06.4 & IC 4732  &  12&5 &   32&1 & 4       &   1&2  & 4    &     3&   & f       &   3&    \\
 010.8+18.0      & M 2- 9   &  39&0 &   36&  & 1       &  46&   & 1    &    17&3  & a       &  17&5   \\
 011.0+05.8      & NGC 6439 &  51&8 &   54&5 & 3, 4    &   4&0 +& 3    &     5&   & f, a    &   4&    \\
 011.0+06.2      & M 2-15   &  19&9 &   13&0 & 1, 5    &   7&4 +& 1, 5 &     5&7  & a       &   7&4   \\
 011.0\,--\,05.1 & M 1-47   &  11&7 &   14&  & 1       &   5&5  & 1    &     4&7  & a       &   5&5   \\
 011.1+11.5      & M 2-13   &  10&1 &   13&3 & 3, 5    &   1&5 +& 3    & $<$ 7&   & i       &   1&5  \\
 011.3\,--\,09.4 & H 2-48   &  32&2 &   68&5 & 1, 5    &   2&0  & 1, 5 &     2&   & d       &   2&    \\
 011.9+04.2      & M 1-32   &  70&5 &   61&  & 1       &   9&0 +& 1    &     7&6  & a       &   9&    \\
 012.5\,--\,09.8 & M 1-62   &  13&6 &   12&8 & 5       &   1&5  & 5    &     3&7  & a       &   1&5   \\
 012.6\,--\,02.7 & M 1-45   &  15&4 &   19&  & 6       &   2&5  & 1    &    ST&   & h       &   2&5   \\
 013.1+04.1      & M 1-33   &  49&2 &   60&  & 1       &   3&8  & 1    &     4&9  & a       &   4&    \\
 014.3\,--\,05.5 & V-V 3-6  &   4&0 &    9&9 & 3, 5    &   1&0 +& 3    &    ST&   & j       &   1&    \\
 015.6\,--\,03.0 & A 44     &  10&0 &   10&  & 1       &    &   &      &    52&   & a       &  52&    \\
 015.9+03.3      & M 1-39   &  58&6 &   99&9 & 2       &   4&0 +& 1    & $<$10&   & a       &   4&    \\
 016.1\,--\,04.7 & M 1-56   &  15&0 &   21&  & 1       &   1&4  & 1    & $<$10&   & a       &   1&5   \\
 016.4\,--\,01.9 & M 1-46   &  78&4 &   81&  & 1       &  12&0 +& 1    &    10&9  & a       &  11&5   \\
 018.0+20.1      & Na 1     &  24&0 &   22&5 & 1, 8    &   8&   & 1    &     5&   & k       &   8&    \\
 018.9+03.6      & M 4- 8   &  10&7 &   19&  & 3       &   1&3 +& 3    &    ST&   & h       &   1&3   \\
 019.2\,--\,02.2 & M 4-10   &  12&2 &   33&  & 3       &   1&2 +& 3    &    ST&   & h       &   1&2   \\
 019.4\,--\,05.3 & M 1-61   &  32&7 &   97&  & 3       &   1&8 +& 3    &    ST&   & h       &   1&8   \\
 019.7+03.2      & M 3-25   &  25&2 &   76&  & 1       &   1&3  & 1    &     3&9  & a       &   1&5   \\
 019.7\,--\,04.5 & M 1-60   &  35&4 &   48&  & 3       &   2&5 +& 3    & $<$10&   & h       &   2&5   \\
 020.7\,--\,05.9 & Sa 1-8   &  13&8 &   11&  & 3       &   5&6 +& 3    &     8&   & i       &   5&6   \\
 020.9\,--\,01.1 & M 1-51   & 249&  &  319&  & 1       &  15&0 +& 1    &     9&1  & a       &  15&    \\
 022.1\,--\,02.4 & M 1-57   &  54&2 &   66&6 & 2       &   8&0 +& 1    &     8&5  & a       &   8&    \\
 023.8\,--\,01.7 & K 3-11   &  15&0 &   17&  & 3       &   3&0 +& 3    &    ST&   & h       &   3&    \\
 023.9\,--\,02.3 & M 1-59   &  90&7 &  108&  & 3       &   4&8 +& 3    &     4&5  & a       &   4&8   \\
 024.1+03.8      & M 2-40   &  49&1 &   33&  & 1       &   5&5  & 1    &     5&0  & a       &   5&5   \\
 024.2+05.9      & M 4- 9   &  45&9 &   42&  & 1       &  47&0 +& 1    &    43&0 +& a       &  45&    \\
 024.8\,--\,02.7 & M 2-46   &  13&6 &   12&0 & 2       &    &   &      &     4&3  & a       &   4&3   \\
 025.4\,--\,04.7 & IC 1295  &  49&5 &   44&  & 1       & 109&0 +& 1    &    92&0 +& f, a, w & 100&    \\
 025.9\,--\,02.1 & Pe 1-15  &   8&7 &    8&5 & 3, 4    &   4&8  & 3    &     5&1 +& a, w    &   5&0   \\
 027.3\,--\,02.1 & Pe 1-18  &   7&1 &   42&  & 1       &   1&2  & 1    &     1&0  & w       &   1&1   \\
 027.4\,--\,03.5 & Vy 1- 4  &  20&5 &   22&  & 3       &   4&   & 3    &     5&3  & w       &   4&5   \\
 027.6+04.2      & M 2-43   &  20&9 &  148&  & 3       &   1&5 +& 3    &     1&9  & w       &   1&5   \\
 027.6\,--\,09.6 & IC 4846  &  40&7 &   43&  & 3       &   2&9 +& 3    &     2&   & d       &   2&9   \\
 028.5+01.6      & M 2-44   &  51&6 &   54&  & 1       &   8&0 +& 1    &     7&5  & a       &   8&    \\
 028.7+02.7      & K 3- 7   &  30&1 &   30&  & 3       &   6&3 +& 3    & $<$ 8&   & h       &   6&3   \\
 031.0+04.1      & K 3- 6   &  10&9 &   55&  & 3       &   0&7  & 3    &    ST&   & h       &   0&7   \\
 031.0\,--\,10.8 & M 3-34   &  30&4 &   29&  & 1       &   8&0 +& 1    &     5&5  & a       &   8&    \\
 031.2+05.9      & K 3- 3   &  36&4 &   34&  & 1       &  11&0 +& 1    &     8&9 +& a, w    &  10&    \\
 032.7\,--\,02.0 & M 1-66   &  46&6 &   59&  & 3       &   2&7 +& 3    & $<$10&   & h       &   2&7   \\
 032.9\,--\,02.8 & K 3-19   &  12&4 &   23&  & 3       &   1&2 +& 3    &    ST&   & h       &   1&2   \\
 033.1\,--\,06.3 & NGC 6772 &  79&3 &   73&  & 1       &  90&0 +& 1    &    90&0 +& f       &  90&    \\
 033.8\,--\,02.6 & NGC 6741 & 132&8 &  153&  & 4       &   6&2  & 7    &     8&0 +& f, a    &   8&    \\
 034.0+02.2      & K 3-13   &  31&0 &   34&  & 3       &   3&7 +& 3    &    ST&   & h       &   3&7   \\
 034.6+11.8      & NGC 6572 & 551&  & 1260&  & 9       &   9&7 +& 9    &    10&8 +& f, a    &  10&    \\
 037.8\,--\,06.3 & NGC 6790 &  53&1 &  242&  & 3, 9, 4 &   1&8 +& 3    &     7&   & f, a    &   1&8   \\
 038.2+12.0      & Cn 3-1   &  60&0 &   64&8 & 4       &   3&7  & 4    &     4&5  & a, l    &   4&5   \\
 039.5\,--\,02.7 & M 2-47   &  39&1 &   45&  & 1       &   6&   & 1    &     6&4  & a       &   6&    \\
 039.8+02.1      & K 3-17   & 320&  &  345&  & 1       &   8&0 +& 1    &    14&8  & a       &    8&    \\
 040.4\,--\,03.1 & K 3-30   &  12&7 &   23&  & 3       &   3&3 +& 3    &    ST&   & h       &    3&3   \\
 041.8\,--\,02.9 & NGC 6781 & 380&  &  310&  & 1       & 130&0 +& 1    &   110&0 +& a, f    &  120&    \\
 042.9\,--\,06.9 & NGC 6807 &   9&3 &   28&5 & 3, 4    &   0&8  & 3    &     2&   & d       &    0&8   \\
 043.1+03.8      & M 1-65   &  21&3 &   23&0 & 2       &   4&0  & 1    &     3&7  & a       &    4&    \\
 045.4\,--\,02.7 & Vy 2- 2  &   6&2 &   40&5 & 10      &   0&6 +& 10   &    14&   & i       &    0&6   \\
 045.9\,--\,01.9 & K 3-33   &   9&1 &   17&  & 3       &   1&1 +& 3    &    ST&   & h       &    1&1   \\
\hline
 \multicolumn{15}{r}{continued on next page}  
\end{tabular}
\end{table*}

\begin{table*}
\centering
\begin{tabular}{p{1.8cm}p{2.0cm}r@{.}p{0.2cm}r@{.}p{0.5cm}p{1.0cm}r@{.}p{0.6cm}p{0.8cm}r@{.}p{0.5cm}p{1.1cm}r@{.}p{0.4cm}}
 \multicolumn{15}{l}{continued from previous page} \\ 
\hline
    \multicolumn{1}{c}{(1)} &
    \multicolumn{1}{c}{(2)} & 
    \multicolumn{2}{c}{(3)} &
    \multicolumn{2}{c}{(4)} & 
    \multicolumn{1}{c}{(5)} & 
    \multicolumn{2}{c}{(6)} &
    \multicolumn{1}{c}{(7)} & 
    \multicolumn{2}{c}{(8)} & 
    \multicolumn{1}{c}{(9)} & 
    \multicolumn{2}{c}{(10)} \\
\hline
 046.3\,--\,03.1 & PB 9       &   33&6 &   40&  & 3    &  9&0 +& 3 &     5&1 +&  w       &    7&    \\
 046.4\,--\,04.1 & NGC 6803   &   69&4 &   88&7 & 4    &  2&4  & 4 &     5&5  &  f, a    &    5&    \\
 048.0\,--\,02.3 & PB 10      &   50&7 &   50&  & 3    &  8&0 +& 3 &    10&   &  i       &    8&    \\
 048.1+01.1      & K 3-29     &   14&2 &   58&  & 3    &  1&   & 3 &    ST&   &  h       &    1&    \\
 050.4\,--\,01.6 & K 4-28     &    5&4 &   19&  & 3    &  0&6 +& 3 &    ST&   &  h       &    0&6   \\
 051.0+03.0      & He 2-430   &   18&6 &   39&  & 3, 4 &  1&7 +& 3 & $<$ 5&   &  i       &    1&7   \\
 051.4+09.6      & Hu 2-1     &   43&0 &  108&  & 3, 4 &  1&8 +& 3 &     2&6  &  l       &    1&8   \\
 051.9\,--\,03.8 & M 1-73     &   48&1 &   43&5 & 1, 4 &  6&0 +& 1 &     5&1  &  a       &    6&    \\
 052.2\,--\,04.0 & M 1-74     &    7&2 &   29&  & 1    &  1&0  & 1 &     0&8  &  w       &    1&    \\
 052.5\,--\,02.9 & Me 1-1     &   36&3 &   45&1 & 2    &  4&7  & 1 & $<$ 8&   &  i       &    4&7   \\
 052.9+02.7      & K 3-31     &   17&3 &   39&  & 3    &  1&5 +& 3 &    ST&   &  h       &    1&5   \\
 053.2\,--\,01.5 & K 3-38     &   29&2 &   29&  & 1, 4 &  4&7  & 1 &     6&   &  a, w    &    5&    \\
 054.4\,--\,02.5 & M 1-72     &    5&7 &   26&  & 3    &  0&7 +& 3 &     0&8  &  w       &    0&7   \\
 055.2+02.8      & He 2-432   &   23&2 &   33&5 & 3, 2 &  2&3 +& 3 &     5&   &  m       &    2&3   \\
 055.3+02.7      & He 1- 1    &   17&3 &   14&  & 1    &  8&0 +& 1 &     8&2 +&  w       &    8&    \\
 055.5\,--\,00.5 & M 1-71     &   84&1 &  204&  & 1    &  2&9  & 1 &     3&8  &  a       &    3&    \\
 055.6+02.1      & He 1- 2    &   16&1 &   15&  & 1    &  4&7  & 1 &     5&   &  a       &    4&7   \\
 056.0+02.0      & K 3-35     &   14&6 &   40&  & 3    &  1&7 +& 3 &    ST&   &  h       &    1&7   \\
 056.4\,--\,00.9 & K 3-42     &   11&6 &   19&  & 3    &  1&2 +& 3 &     3&4  &  a       &    1&2   \\
 057.2\,--\,08.9 & NGC 6879   &   23&5 &   18&  & 1    &  5&   & 1 &     5&   &  a, f    &    5&    \\
 057.9\,--\,01.5 & He 2-447   &   23&6 &   60&  & 3    &  1&2 +& 3 &     5&   &  i       &    1&2   \\
 058.3\,--\,10.9 & IC 4997    &   30&5 &   71&2 & 4    &  1&6  & 4 &     2&   &  f       &    2&    \\
 058.9+01.3      & K 3-40     &   17&2 &   20&  & 3    &  4&   & 3 &    ST&   &  h       &    4&    \\
 059.4+02.3      & K 3-37     &   14&7 &   17&  & 3    &  2&5  & 3 &    ST&   &  h       &    2&5   \\
 060.1\,--\,07.7 & NGC 6886   &   77&9 &   83&9 & 4    &   &   &   &     5&5  &  f, a    &    5&5   \\
 060.5+01.8      & He 2-440   &   26&9 &   43&  & 3    &  2&2 +& 3 &     3&   &  i       &    2&2   \\
 062.4\,--\,00.2 & M 2-48     &   17&0 &   19&  & 1    &  3&1  & 1 &     8&   &  a       &    4&    \\
 063.8\,--\,03.3 & K 3-54     &    5&1 &    7&3 & 3, 4 &  0&8  & 3 &    ST&   &  h       &    0&8   \\
 064.7+05.0      & BD+30 3639 &  245&0 &  630&  & 9    &  7&7 +& 9 &     7&5 +&  f, a    &    7&6   \\
 064.9\,--\,02.1 & K 3-53     &   10&3 &   50&  & 3    &  0&8  & 3 &     1&   &  w       &    0&9   \\
 066.9\,--\,05.2 & PC 24      &   17&6 &   18&  & 1    &  5&   & 1 &     5&   &  a       &    5&    \\
 067.9\,--\,00.2 & K 3-52     &   17&3 &   65&  & 3    &  0&7  & 3 &    ST&   &  h       &    0&7   \\
 068.3\,--\,02.7 & He 2-459   &   14&0 &   64&  & 3    &  1&3 +& 3 &     1&3  &  w       &    1&3   \\
 068.7+01.9      & K 4-41     &   11&8 &   15&  & 3    &  3&0 +& 3 &    ST&   &  h       &    3&    \\
 071.6\,--\,02.3 & M 3-35     &   29&9 &  130&  & 3    &  1&5 +& 3 & $<$ 5&   &  a       &    1&5   \\
 072.1+00.1      & K 3-57     &   47&3 &   60&  & 3    &  6&3 +& 3 &     6&   &  i       &    6&3   \\
 074.5+02.1      & NGC 6881   &   70&9 &  121&  & 3, 4 &  2&6 +& 3 &     3&5  &  a, f    &    2&6   \\
 082.1+07.0      & NGC 6884   &  152&8 &  175&2 & 4    &  3&1  & 4 &     5&3  &  a, f, w &    5&3   \\
 084.9\,--\,03.4 & NGC 7027   & 1373&  & 6130&  & 9    & 11&0 +& 9 &    14&0 +&  f       &   12&5   \\
 086.5\,--\,08.8 & Hu 1-2     &  107&5 &  336&4 & 4    &  1&7  & 4 &     6&5  &  a       &    6&5   \\
 088.7+04.6      & K 3-78     &   15&2 &   17&  & 1, 3 &  3&8 +& 3 &     3&2  &  n       &    3&8   \\
 088.7\,--\,01.6 & NGC 7048   &   46&8 &   37&  & 1    & 70&0 +& 1 &    61&0 +&  f, a    &   65&    \\
 089.8\,--\,05.1 & IC 5117    &   34&5 &  175&  & 3    &  1&5 +& 3 &     1&2  &  l       &    1&5   \\
 093.3\,--\,00.9 & K 3-82     &   37&3 &   30&  & 1    & 25&0 +& 1 &    23&0 +&  w       &   24&    \\
 093.3\,--\,02.4 & M 1-79     &   23&6 &   19&  & 1    & 30&0 +& 1 &    32&5 +&  a       &   31&    \\
 093.4+05.4      & NGC 7008   &  265&  &  217&  & 1    & 95&0 +& 1 &    86&0 +&  f, a    &   90&    \\
 093.5+01.4      & M 1-78     &  372&  &  906&3 & 4    &  3&8  & 4 &     6&4  &  a       &    6&    \\
 095.1\,--\,02.0 & M 2-49     &   28&3 &   35&  & 3    &  2&5 +& 3 &     4&   &  w       &    2&5   \\
 095.2+00.7      & K 3-62     &   59&9 &  115&  & 3    &  2&5 +& 3 &     3&   &  i       &    2&5   \\
 096.3+02.3      & K 3-61     &   16&9 &   14&  & 1    &  6&0 +& 1 &     6&1  &  a       &    6&    \\
 098.2+04.9      & K 3-60     &   28&1 &   43&  & 3    &  1&9 +& 3 &     3&   &  i       &    1&9   \\
 100.0\,--\,08.7 & Me 2-2     &   16&3 &   44&  & 3, 4 &  1&2 +& 3 &     1&6  &  w       &    1&2   \\
 100.6\,--\,05.4 & IC 5217    &   50&8 &   47&5 & 4    &  2&0  & 4 &     6&5 +&  f, a    &    6&5   \\
 101.8+08.7      & A 75       &   21&6 &   17&  & 1    & 57&   & 1 &    56&0 +&  a, f, w &   56&    \\
 103.7+00.4      & M 2-52     &   15&4 &   14&  & 1    & 14&   & 1 &    13&   &  a       &   13&5   \\
 104.1+01.0      & Bl 2- 1    &   22&0 &   54&  & 3    &  1&6 +& 3 &    ST&   &  i       &    1&6   \\
 104.4\,--\,01.6 & M 2-53     &   15&1 &   11&  & 1    & 20&0 +& 1 &    15&   &  a       &   18&    \\
 106.5\,--\,17.6 & NGC 7662   &  646&  &  631&  & 1    & 26&   & 1 &    17&0 +&  f       &   20&    \\
 107.7\,--\,02.2 & M 1-80     &   18&0 &   25&  & 1    &  8&0 +& 1 &     8&   &  a       &    8&    \\
\hline
 \multicolumn{15}{r}{continued on next page}  
\end{tabular}
\end{table*}

\begin{table*}
\centering
\begin{tabular}{p{1.8cm}p{2.0cm}r@{.}p{0.2cm}r@{.}p{0.5cm}p{1.0cm}r@{.}p{0.6cm}p{0.8cm}r@{.}p{0.5cm}p{1.1cm}r@{.}p{0.4cm}}
 \multicolumn{15}{l}{continued from previous page} \\ 
\hline
    \multicolumn{1}{c}{(1)} &
    \multicolumn{1}{c}{(2)} & 
    \multicolumn{2}{c}{(3)} &
    \multicolumn{2}{c}{(4)} & 
    \multicolumn{1}{c}{(5)} & 
    \multicolumn{2}{c}{(6)} &
    \multicolumn{1}{c}{(7)} & 
    \multicolumn{2}{c}{(8)} & 
    \multicolumn{1}{c}{(9)} & 
    \multicolumn{2}{c}{(10)} \\
\hline
 116.2+08.5      & M 2-55   &   26&6 &   19&  & 1    & 40&0 +& 1     &    42&0 +&  a     &   41&    \\
 118.0\,--\,08.6 & Vy 1- 1  &   19&9 &   20&  & 3, 4 &  6&0 +& 3     &     5&2  &  o     &    6&    \\
 119.6\,--\,06.7 & Hu 1-1   &   28&5 &   24&  & 1, 4 & 10&0 +& 1     &     9&   &  w     &   10&    \\
 126.3+02.9      & K 3-90   &   12&8 &   13&9 & 1    & 10&0 +& 1     &     7&1 +&  w     &    8&5   \\
 130.3\,--\,11.7 & M 1- 1   &   14&2 &    8&3 & 1, 4 &  5&   & 1     &     4&3  &  w     &    4&5   \\
 146.7+07.6      & M 4-18   &   19&7 &   20&  & 1, 3 &  3&75+& 1, 3  &     4&   &  k     &    3&75  \\
 147.4\,--\,02.3 & M 1- 4   &   78&2 &   87&  & 3, 4 &  6&0 +& 3     &     4&   &  a     &    6&    \\
 147.8+04.1      & M 2- 2   &   52&5 &   54&  & 1    &  7&   & 1     &     6&   &  f     &    6&5   \\
 159.0\,--\,15.1 & IC  351  &   32&7 &   25&5 & 3, 4 &  7&0 +& 3     &     7&0 +&  f, a  &    7&    \\
 161.2\,--\,14.8 & IC 2003  &   55&3 &   48&  & 3, 4 &  9&0 +& 3     &    10&5 +&  f     &   10&    \\
 165.5\,--\,06.5 & K 3-67   &   33&7 &   42&  & 3    &  2&2 +& 3     &    ST&   &  p     &    2&2   \\
 167.4\,--\,09.1 & K 3-66   &   15&9 &   18&  & 3    &  2&1 +& 3     &    ST&   &  p     &    2&1   \\
 174.2\,--\,14.6 & H 3-29   &   18&6 &   18&  & 1    & 17&   & 1     &    25&   &  w     &   21&    \\
 184.0\,--\,02.1 & M 1- 5   &   42&8 &   69&5 & 3, 4 &  2&3 +& 3     &     2&   &  r     &    2&3   \\
 189.8+07.7      & M 1- 7   &   18&8 &   13&  & 1    & 11&0 +& 1     &     9&   &  a     &   11&    \\
 190.3\,--\,17.7 & J 320    &   30&4 &   22&  & 3, 4 &  7&1 +& 3     &     7&1 +&  a, z  &    7&1   \\
 194.2+02.5      & J 900    &  108&4 &  103&  & 3, 4 &  6&0 +& 3     &     7&4  &  a     &    6&    \\
 198.6\,--\,06.3 & A 12     &   39&3 &   36&  & 1    & 35&0 +& 1     &    37&   &  a     &   35&    \\
 201.7+02.5      & K 4-48   &   12&1 &   14&  & 3    &  2&2  & 3     & $<$ 2&5  &  w     &    2&2   \\
 211.2\,--\,03.5 & M 1- 6   &   54&9 &   86&  & 1    &  2&9  & 1     & $<$ 5&   &  a     &    3&    \\
 212.0+04.3      & M 1- 9   &   23&9 &   27&  & 3    &  2&3 +& 3     &     2&5  &  w     &    2&3   \\
 221.3\,--\,12.3 & IC 2165  &  180&3 &  177&6 & 4    &   &   &       &     9&0 +&  f, a  &    9&    \\
 226.7+05.6      & M 1-16   &   33&0 &   29&5 & 3, 4 &  3&6 +& 3     &     3&   &  a     &    3&6   \\
 228.8+05.3      & M 1-17   &   18&5 &   17&  & 3, 4 &  2&5 +& 3     &     3&   &  a     &    2&5   \\
 231.8+04.1      & NGC 2438 &   80&1 &   67&  & 1    & 80&0 +& 1     &    65&0 +&  f, a  &   72&    \\
 232.4\,--\,01.8 & M 1-13   &   19&2 &   13&6 & 4    &   &   &       &    10&   &  a     &   10&    \\
 232.8\,--\,04.7 & M 1-11   &   26&1 &  113&  & 3    &  2&2 +& 3     &    ST&   &  d     &    2&2   \\
 234.8+02.4      & NGC 2440 &  391&  &  370&  & 1    & 18&0 +& 1     &    16&   &  a     &   17&    \\
 234.9\,--\,01.4 & M 1-14   &   59&4 &   60&  & 3    &  4&7 +& 3     &    ST&   &  h     &    4&7   \\
 235.3\,--\,03.9 & M 1-12   &   22&2 &   41&  & 3    &  1&8 +& 3     &    ST&   &  h     &    1&8   \\
 239.6+13.9      & NGC 2610 &   30&7 &   30&  & 1    & 49&0 +& 1     &    38&0 +&  f, a  &   43&    \\
 242.6\,--\,11.6 & M 3- 1   &   26&0 &   24&  & 1    & 11&   & 1     &    11&2  &  a     &   11&2   \\
 245.4+01.6      & M 3- 5   &   10&9 &   10&  & 1    &  7&0 +& 1     &     6&8  &  a     &    7&    \\
 248.8\,--\,08.5 & M 4- 2   &   24&7 &   19&  & 1    &  6&0 +& 1     &     6&2  &  a     &    6&    \\
 253.9+05.7      & M 3- 6   &   77&8 &   75&  & 1    & 11&0 +& 1     &     8&2  &  a     &   11&    \\
 331.3+16.8      & NGC 5873 &   45&9 &   39&7 & 4    &  1&1  & 4     &     7&   &  d     &    7&    \\
 341.6+13.7      & NGC 6026 &   15&8 &   22&  & 1    & 42&   & 1     &    40&0 +&  f     &   40&    \\
 342.1+27.5      & Me 2-1   &   33&9 &   26&  & 3, 4 &  7&0 +& 3     &     7&   &  a     &    7&    \\
 347.4+05.8      & H 1- 2   &   14&6 &   62&  & 1    &  0&8  & 1     &    ST&   &  i     &    1&    \\
 348.0+06.3      & RPZM 2-1 &   15&6 &   10&2 & 13   &   &   &       &     9&   &  s     &    9&    \\
 349.5+01.0      & NGC 6302 & 1908&  & 2800&  & 11   &   &   &       &    45&   &  a     &   45&    \\
 349.8+04.4      & M 2- 4   &   24&9 &   32&  & 6    &  2&   & 1     & $<$ 5&   &  a     &    2&    \\
 350.9+04.4      & H 2- 1   &   51&9 &   61&  & 1    &  2&2  & 1     &     2&6  &  u     &    2&4   \\
 351.1+04.8      & M 1-19   &   23&4 &   26&  & 1    &  2&6  & 1     &     3&0  &  u     &    2&8   \\
 351.2+05.2      & M 2- 5   &   14&4 &   12&  & 6    &  5&   & 1     &     5&1  &  a     &    5&    \\
 352.1+05.1      & M 2- 8   &   16&2 &   18&  & 1    &  3&7  & 1     &     4&2  &  a     &    4&    \\
 353.3+06.3      & M 2- 6   &   15&2 &   17&  & 6    &  1&6  & 1     &     2&2  &  u     &    2&    \\
 353.5\,--\,04.9 & H 1-36   &   13&5 &   50&  & 3    &  0&8  & 3     &    ST&   &  i     &    0&8   \\
 355.1\,--\,02.9 & H 1-31   &    6&3 &   16&  & 6    &  0&7  & 1     & $<$ 5&   &  a     &    0&7   \\
 355.4\,--\,02.4 & M 3-14   &   23&0 &   30&  & 1    &  2&8  & 1     &     5&2  &  u     &    4&    \\
 355.7\,--\,03.0 & H 1-33   &   16&2 &   12&  & 1    &  2&7  & 1     &    ST&   &  h     &    2&7   \\
 355.7\,--\,03.5 & H 1-35   &   18&8 &   72&  & 1    &  1&1  & 1     &     2&0  &  m, u  &    2&0   \\
 355.9+02.7      & Th 3-10  &   21&6 &   29&5 & 1    &  2&   & 1     &    ST&   &  h     &    2&    \\
 355.9\,--\,04.2 & M 1-30   &   23&6 &   31&  & 6    &  3&5  & 1     & $<$ 5&   &  a     &    3&5   \\
 356.2\,--\,04.4 & Cn 2-1   &   25&5 &   49&  & 1    &  1&7  & 1     &     2&4  &  m, u  &    2&2   \\
 356.5\,--\,02.3 & M 1-27   &   65&4 &   63&  & 1    &  8&0 +& 1     &     5&3  &  a     &    8&    \\
 356.5\,--\,03.9 & H 1-39   &   11&1 &   13&  & 1    &  1&7  & 1     & $<$ 5&   &  a     &    1&7   \\
 356.7\,--\,04.8 & H 1-41   &   16&9 &   12&  & 1    &   &   &       &     9&6  &  a     &    9&6   \\
 356.9+04.5      & M 2-11   &   17&7 &   22&  & 1    &  2&7  & 1     &     1&7  &  u     &    2&2   \\
\hline
 \multicolumn{15}{r}{continued on next page}  
\end{tabular}
\end{table*}

\begin{table*}
\centering
\begin{tabular}{p{1.8cm}p{2.0cm}r@{.}p{0.2cm}r@{.}p{0.5cm}p{1.0cm}r@{.}p{0.6cm}p{0.8cm}r@{.}p{0.5cm}p{1.1cm}r@{.}p{0.4cm}}
 \multicolumn{15}{l}{continued from previous page} \\ 
\hline
    \multicolumn{1}{c}{(1)} &
    \multicolumn{1}{c}{(2)} & 
    \multicolumn{2}{c}{(3)} &
    \multicolumn{2}{c}{(4)} & 
    \multicolumn{1}{c}{(5)} & 
    \multicolumn{2}{c}{(6)} &
    \multicolumn{1}{c}{(7)} & 
    \multicolumn{2}{c}{(8)} & 
    \multicolumn{1}{c}{(9)} & 
    \multicolumn{2}{c}{(10)} \\
\hline
 357.1+03.6      & M 3- 7  &  33&3 &  28&  & 1     &  4&7  & 1     &     5&8  &  a       &    5&    \\
 357.2+07.4      & M 4- 3  &  12&5 &  28&  & 6     &  1&4  & 1     &     1&7  &  u       &    1&6   \\
 357.4\,--\,03.2 & M 2-16  &  20&7 &  25&  & 1     &  2&2  & 1     &     2&7  &  u       &    2&5   \\
 357.4\,--\,03.5 & M 2-18  &  11&4 &  17&  & 1     &  1&5  & 1     & $<$ 5&   &  a       &    1&5   \\
 357.6+01.7      & H 1-23  &  24&8 &  35&  & 6     &  2&5  & 1     &     2&8  &  a       &    2&6   \\
 357.6+02.6      & H 1-18  &  17&9 &  32&6 & 12, 6 &  1&5  & 1     &    ST&   &  h       &    1&5   \\
 358.2+03.5      & H 2-10  &  13&4 &  20&  & 1     &  2&0 +& 1     & $<$ 5&   &  a       &    2&    \\
 358.2+03.6      & M 3-10  &  35&3 &  46&5 & 12, 8 &  3&   & 1     &     3&3  &  a       &    3&1   \\
 358.2+04.2      & M 3- 8  &  19&3 &  20&  & 1     &  2&9  & 1     &     3&5  &  u       &    3&2   \\
 358.3+03.0      & H 1-17  &   6&9 &  35&  & 1     &  0&9  & 1     & $<$ 5&   &  a       &    0&9   \\
 358.3\,--\,02.5 & Al 2- O &  38&3 &  36&8 & 13    &  2&4  & 13    &      &   &          &    2&4   \\
 358.5+05.4      & M 3-39  & 280&  & 280&  & 1     & 20&0 +& 1     &    18&   &  a       &   19&    \\
 358.5\,--\,02.5 & M 4- 7  &  40&0 &  33&  & 1     &  5&7  & 1     &     7&   &  i       &    5&7   \\
 358.5\,--\,04.2 & H 1-46  &  19&2 &  43&  & 1     &  1&2  & 1     &    ST&   &  i       &    1&2   \\
 358.6+01.8      & M 4- 6  &  26&8 &  42&  & 6     &  2&   & 1     &    ST&   &  h       &    2&    \\
 358.7+05.2      & M 3-40  &  16&3 &  17&  & 1     &  2&5  & 1     &    ST&   &  h       &    2&5   \\
 358.7\,--\,05.2 & H 1-50  &  20&8 &  31&  & 1     &  1&4  & 1     & $<$10&   &  a       &    1&4   \\
 358.8+03.0      & Th 3-26 &  10&7 &  10&  & 1     &  6&5  & 1     &     6&   &  d       &    6&5   \\
 358.9+03.2      & H 1-20  &  27&3 &  32&  & 1     &  3&3 +& 1     &     4&0  &  a       &    3&3   \\
 358.9+03.4      & H 1-19  &  12&4 &  26&  & 1     &  1&4  & 1     &    ST&   &  h       &    1&4   \\
 358.9\,--\,00.7 & M 1-26  &  66&8 & 400&  & 3, 4  &  3&2 +& 3     &     4&1  &  a       &    3&2   \\
 359.1\,--\,01.7 & M 1-29  &  91&9 &  97&  & 1     &  7&0 +& 1     &     7&3  &  a, g, b &    7&    \\
 359.2+01.2      & 19W32   &  13&9 &  21&3 & 13    &   &   &       &    24&   &  t       &   24&    \\
 359.3\,--\,00.9 & Hb 5    & 180&  & 471&  & 7     &  3&4  & 7     &    15&   &  a       &    3&4   \\
 359.3\,--\,01.8 & M 3-44  &  25&3 &  35&  & 6     &  4&   & 1     &     4&4  &  a, g    &    4&2   \\
 359.3\,--\,03.1 & M 3-17  &  11&4 &  12&  & 1     &  2&9 +& 1     & $<$ 5&   &  a       &    2&9  \\
 359.7\,--\,01.8 & M 3-45  &  21&5 &  24&3 & 13    &  3&3  & 13    &     5&8  &  a, g, b &    3&5   \\
 359.7\,--\,02.6 & H 1-40  &   8&5 &  31&  & 6     &  1&3  & 1     &     3&8  &  g, b, e &    3&0   \\
 359.8+03.7      & Th 3-25 &  15&6 &  18&  & 1     &  1&8  & 1     &     2&2  &  u       &    2&    \\
 359.9+05.1      & M 3- 9  &  37&6 &  35&  & 1     & 17&   & 1     &    15&   &  a       &   16&    \\
 359.9\,--\,04.5 & M 2-27  &  22&8 &  50&  & 6     &  2&5  & 1     &     4&8  &  b, e    &    3&8   \\
\hline \\[0.0cm]
\end{tabular}
\begin{flushleft}
References: \\[0.15cm]
F(5GHz) \& radio diameter ref.: \\
1 - Zijlstra et al. 1989;
2 - Pottasch \& Zijlstra 1994;
3 - Aaquist \& Kwok 1990;
4 - Isaacman 1984;
5 - Ratag \& Pottasch 1991;
6 - Gathier et al. 1983;
7 - Phillips \& Mampaso 1988;
8 - Kwok et al. 1981;
9 - Basard \& Daub 1987;
10 - Seaquist \& Davis 1983;
11 - Rodriguez et al. 1985;
12 - Pottasch et al. 1988;
13 - Ratag et al.1990;
14 - Zijlstra \& Pottasch 1989.\\[0.15cm]
Optical diameter ref.: \\
a - Perek \& Kohoutek 1967;
b - Kinman et al. 1988;
c - Holmberg et al. 1977;
d - Cahn \& Kaler 1971;
e - Dopita et al. 1990;
f - Chu et al. 1987;
g - Moreno et al. 1988;
h - Schohn 1990;
i - Acker et al. 1991;
j - Weinberger 1977;
k - Shaw 1985;
l - Kohoutek \& Martin 1981;
m - Cahn \& Kaler 1988;
n - Kohoutek 1972;
o - Shaw \& Kaler 1985;
p - Kohoutek 1969;
r - Lauberts 1982;
s - Manchado et al. 1989;
t - Isaacman et al. 1980;
u - Bedding \& Zijlstra 1994;
w - Manchado et al. 1996;
z - Balick 1987.
\end{flushleft}

\end{table*}

\end{document}